\documentclass[%
 reprint,
superscriptaddress,
 amsmath,amssymb,
 aps,
 pra,
]{revtex4-1}
\pdfoutput=1
\usepackage[margin=0.8in]{geometry}                
\usepackage[parfill]{parskip}    
\usepackage{epstopdf}
\usepackage{graphicx}
\usepackage{subcaption}
\usepackage[export]{adjustbox}
\usepackage{setspace}
\graphicspath{ {./Images/} }
\usepackage{lettrine}
\usepackage{type1cm}
\usepackage[separate-uncertainty = true,multi-part-units=single, bracket-numbers = false, scientific-notation=true, round-precision=2]{siunitx}
\usepackage{array}
\usepackage{float}								
\usepackage{setspace}
\usepackage{physics}
\usepackage{braket}
\usepackage{mathtools}
\usepackage[]{hyperref}
\hypersetup{breaklinks=true}
\usepackage{qcircuit}
\usepackage{dsfont}
\usepackage{mathtools}
\graphicspath{{./images/}}
\def\ketbra#1#2{{\vert#1\rangle\!\langle#2\vert}}
\usepackage{xcolor}
\usepackage[normalem]{ulem}

\renewcommand{\thefootnote}{\fnsymbol{footnote}}

\begin{document}

\title{Quantum State Discrimination Using Noisy Quantum Neural Networks}

\author{Andrew Patterson}%
    \email{andrew.patterson.10@ucl.ac.uk}
\affiliation{Dept.\ of Physics and Astronomy, University College London, \\
            Gower Street, London,\\
            WC1E 6BT, United Kingdom}%
\author{Hongxiang Chen}
\affiliation{Rahko Ltd.\, \\ 
             Finsbury Park, N4 3JP, \\
             United Kingdom}%
\affiliation{Dept.\ of Computer Science, University College London, \\
            Gower Street, London,\\
            WC1E 6BT, United Kingdom}%
\author{Leonard Wossnig}
\affiliation{Rahko Ltd.\, \\ 
             Finsbury Park, N4 3JP, \\
             United Kingdom}%
\affiliation{Dept.\ of Computer Science, University College London, \\
            Gower Street, London,\\
            WC1E 6BT, United Kingdom}%
\author{Simone Severini}
\affiliation{Dept.\ of Computer Science, University College London, \\
            Gower Street, London,\\
            WC1E 6BT, United Kingdom}%
\author{Dan Browne}
\affiliation{Dept.\ of Physics and Astronomy, University College London, \\
            Gower Street, London,\\
            WC1E 6BT, United Kingdom}%
\author{Ivan Rungger}%
    \email{ivan.rungger@npl.co.uk}
\affiliation{National Physical Laboratory, \\ 
            Teddington, TW11 0LW, \\
            United Kingdom}%
\affiliation{Dept.\ of Physics and Astronomy, University College London, \\
            Gower Street, London,\\
            WC1E 6BT, United Kingdom}%
\date{\today}

\begin{abstract}
	Near-term quantum computers are noisy, and therefore must run algorithms with a low circuit depth and qubit count.
	Here we investigate how noise affects a Quantum Neural Network (QNN) for state discrimination, which is applicable on near-term quantum devices as it fulfils the above criteria.
	We find that for the required gradient calculation on a noisy device a quantum circuit with a large number of parameters is disadvantageous. By introducing a new smaller circuit ansatz we overcome this limitation, and find that the QNN performs well at noise levels of current quantum hardware.
	We present a model showing that the main effect of the noise is to increase the overlap between the states as circuit gates are applied, hence making discrimination more difficult.
	Our findings demonstrate that noisy quantum computers can be used for state discrimination and other applications, such as classifiers of the output of quantum generative adversarial networks.
	\vspace{10mm}
\end{abstract}

\maketitle

\footnotetext[0]{Corresponding authors:
\\$^*$\href{mailto:andrew.patterson.10@ucl.ac.uk}{andrew.patterson.10@ucl.ac.uk}, 
\\$^\dagger$\href{mailto:ivan.rungger@npl.co.uk}{ivan.rungger@npl.co.uk}}
\renewcommand{\thefootnote}{\arabic{footnote}}
	
\section{Introduction}\label{sec:intro}
Quantum state discrimination is important in many emerging quantum technologies: quantum cryptography \cite{Bennett1992}, entanglement concentration \cite{Chefles2000}, quantum cloning \cite{Duan1998}, and quantum metrology and sensing \cite{Giovannetti2011, Lloyd2008}.
Quantum circuits trained for classification could also be used in quantum machine learning problems as a classifier of quantum data. They could classify the output of other quantum circuits, e.g. the output of a quantum generative adversarial network (GAN) \cite{Benedetti2019}.
Current quantum computing devices are subject to non-negligible amounts of noise \cite{IBM2019, Hong2019, Chen2014}, and therefore algorithm design for devices in the near future must take this into account.
Here we present an extension to noisy devices of the approach for quantum state discrimination outlined in Ref. \cite{Chen2018}, a quantum analogue of a neural network used for state discrimination.
In Ref. \cite{Chen2018} simulations of shallow quantum circuits were trained to find the optimal Positive Operator Valued Measure (POVM), or measurement, to distinguish between two families of non-orthogonal quantum states.
Given an input state chosen randomly from one of the families, the output of the network should indicate which family the input was chosen from.
To do this the network is trained on a set of labelled data, performing supervised learning \cite{Goodfellow2016}.
The ideal POVM was learned via a classical optimiser using a gradient descent algorithm on the quantum parameters, which correspond to the rotation gates in the quantum circuit.
This POVM is distinct from the error minimising POVMs, as it also attempts to minimise the occurrence of inconclusive results.

There are similarities between a classical unitary neural network \cite{Arjovsky2015} and this algorithm.
It contains a layer of symmetric fully connected neurons, followed by arbitrary numbers of non-linear layers, or dropout layers.
The non-linearity in the quantum network is introduced by measurement of some of the qubits.

In this work we extend the simulations done previously from pure vector states to simulations of states represented as density matrices, so that we can model noise in the quantum device.
We also simulate calculation of the parameter gradients on the quantum device, which would also be subject to noise in a real machine.
We find that with these extensions including the effect of noise the previous algorithm proposed for noiseless systems no longer performs optimally.
To recover performance we reduce the number of trainable parameters through consideration of the circuit structure.

This paper is structured as follows: we begin by outlining the theory of state discrimination and the QNN. We then discuss the simulation methods, gradient calculation and measurement. In Section~\ref{sec:results} we present the results, including the effect of reducing the number of parameters and the effects of noise on training the circuits.

\section{Methods}\label{sec:methods}
\subsection{Quantum State Discrimination} \label{sec:methods_qsd}
We wish to discriminate a two-qubit input state, $\ket{\psi_{in}}$, which in general can be represented as a normalised vector with 4 complex components.
The state is chosen randomly from two families of states, labelled $a$ and $b$:
\begin{equation}\label{eqn:states_a}
    \ket{\psi_{in,a}} = (\sqrt{1-a^2}, 0, a, 0),
\end{equation}
\begin{equation}\label{eqn:states_b}
    \ket{\psi_{in,b}} = (0, \pm \frac{1}{\sqrt{2}}, \frac{1}{\sqrt{2}}, 0),
\end{equation}

Note however that our results are expected to be applicable also to other families of states,  since the methods presented here are based on variational algorithms, which can be formulated for any target state.

To simulate the effect of noise we use density matrices to represent quantum states:
\begin{equation}
\rho_{a(b)} = \ketbra{\psi_{a(b)}}{\psi_{a(b)}}.
\end{equation}

The value of $a$ is drawn from a probability distribution $a \in (0, 1]$ which is characterised by its mean, $\mu_a$ and standard deviation, $\sigma_a$.
Note that the $b$ states can take two values, depending on the sign in the second vector element; these are selected randomly, with equal probability.
We set the probability that an $a$ state appears in the data set to $p_a = 1 / 3$, the probability that a $b$ state with positive sign appears to $p_{p+}=1/3$, and the probability that a $b$ state with a negative sign appears to $p_{p-}=1/3$.

Primarily these states are chosen to reproduce the work in \cite{Chen2018} and in \cite{Mohseni2004}, where the state discrimination was performed in a laboratory.
Secondly, these states are non-orthogonal and therefore cannot be distinguished perfectly without some probability of erroneous or inconclusive outcomes, making the problem harder for the algorithm.
It is therefore an ideal case to verify the method, since the level of non-orthogonality can be tuned by choosing the value of $a$ in Eq. (\ref{eqn:states_a}), with a value closer to $1$ being more difficult to discriminate.

\subsection{The Quantum Neural Network}
\label{sec:methods_QNN}
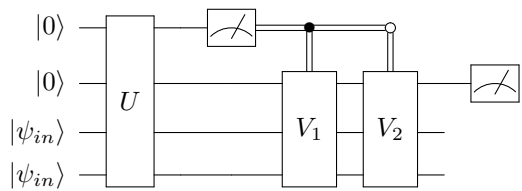
\begin{figure}[ht]
	\centering
	\mbox{
		\Qcircuit @C=1em @R=.7em {
		\lstick{\ket{0}}      & \multigate{3}{U} & \qw & \meter & \control{1} \cw         & \controlo{1} \cw                       \\
		\lstick{\ket{0}}      & \ghost{U}        & \qw & \qw    & \multigate{2}{V_1} \cwx & \multigate{2}{V_2} \cwx & \qw & \meter \\
		\lstick{\ket{\psi_{in}}} & \ghost{U}        & \qw & \qw    & \ghost{V_1}             & \ghost{V_2}             & \qw          \\
		\lstick{\ket{\psi_{in}}} & \ghost{U}        & \qw & \qw    & \ghost{V_1}             & \ghost{V_2}             & \qw    
		}
	}
	\caption{\label{fig:gen_circuit} The general form of the quantum circuits used in this work. The input state is on the bottom two qubits, and measuring the first qubit introduces a non-linear dropout layer. The sub-circuits $U$, $V_{1,2}$ are shown in Figure~\ref{fig:U_V_circuits}.}
\end{figure}

In a classical neural network some nodes are discarded during training, which is called dropout and stops the network from over-fitting \cite{Srivastava2014}.
We can also think of dropout as the introduction of non-linearity into the network. A network with dropout cannot be represented by any smaller, linear network, whereas a many-layered linear network can always be reduced to a single linear layer.
Quantum evolution is unitary and linear, so if we wish to introduce non-linearity into a quantum neural network we need to include a measurement.
Figure~\ref{fig:gen_circuit} shows the structure of the QNN, where the choice of the second step of the circuit, $V_{1, 2}$ is conditioned on the outcome of a measurement on the first qubit.
The measurement results are then used as the output of the neural network \cite{Chen2018}.

There are two non-orthogonal states to discriminate, so if we wish to have a network that can be trained to not commit any errors, we must allow for it to produce an inconclusive result \cite{Mohseni2004}.
This allows the network to give a `don't know' result as opposed to an erroneous one.
Therefore we have a minimum of three outputs, necessitating two measurement qubits.

The output of the network is determined by the measurement outcome.
As we begin in a random configuration and are training the system, we can arbitrarily select which label a measurement outcome corresponds to:
\begin{equation}
    \{\ket{00}: a, \: \ket{01}: b, \: \ket{10}: a, \: \ket{11}: \text{inconclusive}\}.
\label{eqn:measurement_labels}
\end{equation}
The choice of unbalanced labels may have an effect upon the outcome. For a random measurement outcome the probability to guess the right state is 1/2 for $a$ and 1/4 for $b$. 
We partly mitigate this bias setting the probability of $a$ and $b$ states to appear as input to the values specified in the previous subsection, namely $p_a = 1 / 3$ and $p_b =p_{b+}+p_{b-}= 2 / 3$.

This results in the probability of correctly guessing the input state for a fully random measurement outcome to be $1/2 p_a+1/4 p_b=1/3$, and correspondingly the probability for an incorrect guess is equal to $2/3$.
In general one might adapt the assignment of measurement outcomes to labels according to the considered specific task.

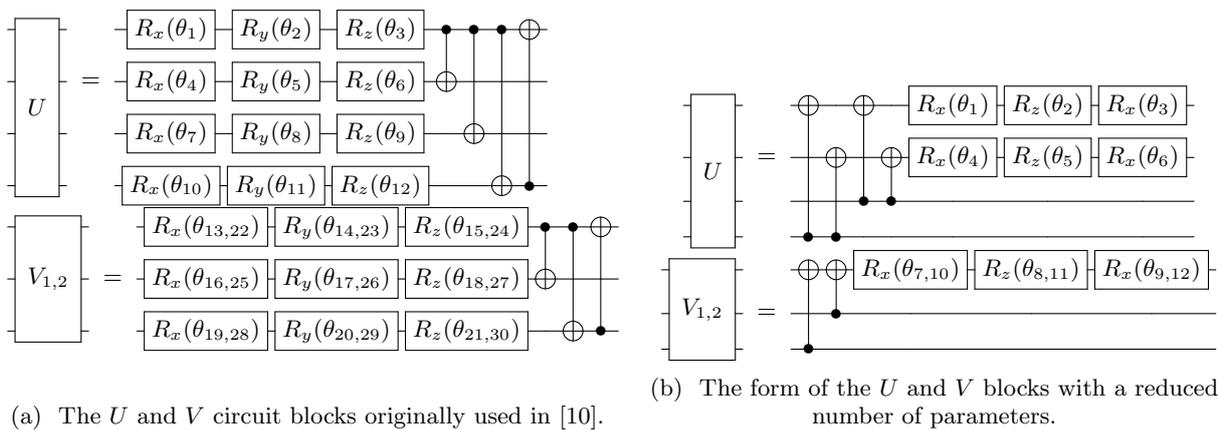
\begin{figure*}[ht]
	\begin{subfigure}[b]{0.48\textwidth}
		\begin{flushleft}
		\mbox{
			\small
			\Qcircuit @C=.3em @R=.55em{
				& \multigate{3}{U} & \qw &                                          &  & \gate{R_x(\theta_1)}  & \gate{R_y(\theta_2)}  & \gate{R_z(\theta_3)} & \ctrl{1}  & \ctrl{2} & \ctrl{3} & \targ     & \qw \\   
				& \ghost{U}        & \qw & \push{\rule{.01em}{0em}=\rule{.01em}{0em}} &  & \gate{R_x(\theta_4)}  & \gate{R_y(\theta_5)}  & \gate{R_z(\theta_6)} & \targ     & \qw      & \qw      & \qw       & \qw \\
				& \ghost{U}        & \qw &                                          &  & \gate{R_x(\theta_7)}  & \gate{R_y(\theta_8)}  & \gate{R_z(\theta_9)} & \qw       & \targ    & \qw      & \qw       & \qw \\
				& \ghost{U}        & \qw &                                          &  & \gate{R_x(\theta_{10})} & \gate{R_y(\theta_{11})} & \gate{R_z(\theta_{12})} & \qw       & \qw      & \targ    & \ctrl{-3} & \qw 
			}
		}
		\vspace{5mm}
		\mbox{
			\small
			\Qcircuit @C=.3em @R=.55em{
				& \multigate{2}{V_{1,2}} & \qw &                                          &  & \gate{R_x(\theta_{13,22})} & \gate{R_y(\theta_{14,23})} & \gate{R_z(\theta_{15,24})} & \ctrl{1}  & \ctrl{2} & \targ     & \qw \\   
				& \ghost{V_{1,2}}        & \qw & \push{\rule{.01em}{0em}=\rule{.01em}{0em}} &  & \gate{R_x(\theta_{16,25})} & \gate{R_y(\theta_{17,26})} & \gate{R_z(\theta_{18,27})} & \targ     & \qw      & \qw       & \qw \\
				& \ghost{V_{1,2}}        & \qw &                                          &  & \gate{R_x(\theta_{19,28})} & \gate{R_y(\theta_{20,29})} & \gate{R_z(\theta_{21,30})} & \qw       & \targ    & \ctrl{-2} & \qw 
			}
		}
		\caption{\label{fig:U_V_long} The $U$ and $V$ circuit blocks originally used in \cite{Chen2018}.}
		\end{flushleft}
	\end{subfigure}
	\begin{subfigure}[b]{0.48\textwidth}
		\mbox{
			\small
			\Qcircuit @C=.3em @R=.55em{
				& \multigate{3}{U} & \qw &                                          &  & \targ     & \qw       & \targ     & \qw       & \gate{R_x(\theta_1)} & \gate{R_z(\theta_2)} & \gate{R_x(\theta_3)} & \qw \\   
				& \ghost{U}        & \qw & \push{\rule{.01em}{0em}=\rule{.01em}{0em}} &  & \qw       & \targ     & \qw       & \targ     & \gate{R_x(\theta_4)} & \gate{R_z(\theta_5)} & \gate{R_x(\theta_6)} & \qw \\
				& \ghost{U}        & \qw &                                          &  & \qw       & \qw       & \ctrl{-2} & \ctrl{-1} & \qw        & \qw        & \qw        & \qw \\
				& \ghost{U}        & \qw &                                          &  & \ctrl{-3} & \ctrl{-2} & \qw       & \qw       & \qw        & \qw        & \qw        & \qw 
			}
		}
		\hfill
		\hfill
		\mbox{
			\small
			\Qcircuit @C=.3em @R=.55em{
				& \multigate{2}{V_{1,2}} & \qw &                                          &  & \targ     & \targ     & \gate{R_x(\theta_{7,10})} & \gate{R_z(\theta_{8,11})} & \gate{R_x(\theta_{9,12})} & \qw \\   
				& \ghost{V_{1,2}}        & \qw & \push{\rule{.01em}{0em}=\rule{.01em}{0em}} &  & \qw       & \ctrl{-1} & \qw        & \qw        & \qw        & \qw \\
				& \ghost{V_{1,2}}        & \qw &                                          &  & \ctrl{-2} & \qw       & \qw        & \qw        & \qw        & \qw 
			}
		}
		\caption{\label{fig:U_V_short} The form of the $U$ and $V$ blocks with a reduced number of parameters.}
	\end{subfigure}
	\caption{\label{fig:U_V_circuits}The circuits showing the trainable parameters, which are used in this work. Comparison of results obtained for the circuits \ref{fig:U_V_short} and \ref{fig:U_V_long} is made in Section~\ref{sec:wo_noise}.}
\end{figure*}

The structure of the $U$ and $V_{1,2}$ circuit blocks is given in Figure~\ref{fig:U_V_circuits}, where \ref{fig:U_V_long} shows the same circuits used in \cite{Chen2018} and \ref{fig:U_V_short} shows the reduced circuits introduced here, which we will discuss in more detail below.
These circuits are small and have low-depth, so that they can be ran on a quantum computer which supports measurement as the circuit is running and classical feedback.
This requires fast measurement and fast classical processing which is not possible in many current systems, but has been achieved in an ion-trap device \cite{Negnevitsky2018}, meaning this algorithm could run on a current device.

The state discrimination task is then as follows: given a set of randomly selected and labelled input states, the classical optimiser must optimise the rotation angles $\theta_{1..n}$ of the quantum circuit to maximise the likelihood of a correct determination of the state. In our specific case it has to determine whether an input state is an $a$ or a $b$ state. Note that only these states are allowed as input states during both training and testing of the circuit.
A correct determination is found when the measurement output of the quantum circuit is equal to the corresponding input state label as defined in Equation ~\ref{eqn:measurement_labels}.

\subsection{Optimisation}\label{sec:methods_optimisation}
Since the input states are initially labelled, the task for the classical optimiser is a supervised learning task \cite{Goodfellow2016}.
The optimiser used in this experiment is Adam \cite{Kingma2014}, which has been found to work well in a number of quantum variational algorithms \cite{Chen2018, Gokhale2019, Sweke2019, Hamilton2019, Liu2019}.
It has also  been shown classically that Adam deals well with noisy gradients \cite{Neelakantan2015}, which will be the output of our noisy quantum computer.
This is possible since Adam uses the concept of momentum, where the gradients of past steps contribute to the current step.
Other optimisers such as RotoSolve \cite{Ostaszewski2019} have been proposed, and a comparison of performance can be made in future work.

Noisy gradients are a feature of the work here: as gradient calculation must be performed on the noisy quantum device, we expect that the output gradients will be noisy.
We also expect that there will be non-optimal local minima in our loss landscape, as this is also a feature of the loss function in the noiseless case \cite{Chen2018}. Finally we also expect that the loss landscape may feature `barren plateaus', as these have been shown to be a feature of quantum optimisation problems \cite{McClean2018}.
This further motivates the choice of a gradient-based optimiser such as Adam.

We define the function to minimise, the cost function, as
\begin{equation}\label{eqn:loss}
    C = \alpha_{\textrm{err}} \textrm{P}_{\textrm{err}} + \alpha_{\textrm{inc}} \textrm{P}_{\textrm{inc}},
\end{equation}
where the positive real numbers $\alpha_{\textrm{err}}, \alpha_{\textrm{inc}}$ are the cost parameters used to bias the network towards minimising errors or inconclusive results ($\textrm{P}_{\textrm{err}}, \textrm{P}_{\textrm{inc}}$ are defined below).
If for example we require the network to produce fewer errors, we can do this at the cost of recording more inconclusive results by increasing the value of $\alpha_{\textrm{err}}$ relative to the value of $\alpha_{\textrm{inc}}$.
We discuss the effect of changing the cost parameters in Section~\ref{sec:wo_noise}.

The measurement probabilities of a state, $\rho$, for a generalised measurement, $M = \ketbra{\phi}{\phi}$,  are given by:
\begin{equation}
    \langle \rho \rangle = \text{Tr}(\ketbra{\phi}{\phi}\rho),
\end{equation}
and the  quantum state after measurement is given by
\begin{equation}
    \rho_{\textrm{measured}} = \frac{\ketbra{\phi}{\phi} \rho \ketbra{\phi}{\phi} }{\text{Tr}(\ketbra{\phi}{\phi}\rho)}.
\end{equation}
Using this we can find the probability of an erroneous or inconclusive measurement
\begin{equation}
    \textrm{P}_{\textrm{err}} = \sum_{\rho_{i} \in b} (\langle \rho_{i} \rangle_{00} + \langle \rho_{i} \rangle_{10}) + \sum_{\rho_{i} \in a} \langle \rho_{i} \rangle_{01} ,
\end{equation}
\begin{equation}
    \textrm{P}_{\textrm{inc}} = \sum_{\rho_i \in a,b}  \langle \rho_{i} \rangle_{11} ,
\end{equation}
where $\rho_i$ is the input state, and $\langle \rho_i \rangle_{jk}$ refers to the probability of obtaining a measurement of $\ket{jk}$ from the circuit.

Discrimination of these states, without the use of a variational algorithm, has been shown in the laboratory to reach the theoretical best success probability, $\textrm{P}_{\textrm{suc}}$ of $0.833$ for $\mu_a = 0.25, \sigma_a = 0.01$ \cite{Mohseni2004}.
This is a minimum loss, $L=1-\textrm{P}_{\textrm{suc}}=\textrm{P}_{\textrm{err}} +\textrm{P}_{\textrm{inc}}$, of $0.166$.
For the equal probability case, $P(\ket{00}) = P(\ket{01}) = P(\ket{10}) = P(\ket{11}) = 0.25$, the success rate is $0.385$, this translates into a  loss of $0.635$.
This gives us lower and upper expected bounds to compare our results for the loss to.

\subsection{Gradient Calculation}\label{sec:methods_gradients}
Unlike gradient-free optimisers (such as Nelder-Mead \cite{Nelder1965}) the Adam optimiser requires the calculation of parameter derivatives ($\frac{\partial \langle C \rangle}{\partial \theta_{0..n}}$).
In the previous work this was done using the forward differences formula \cite{Chen2018}, which requires direct access to the components of the wavefunction.
In a real quantum computer this is difficult to achieve, and hence here we use a more practical approach.
Calculation of the gradients of quantum parameters has received attention recently \cite{Mitarai2018, Schuld2018, Parrish2019} due to the introduction of variational methods such as the variational quantum eigensolver (VQE) \cite{Peruzzo2014}.
The gradient of the loss function with respect to a parameter $\theta_i$ is calculated by the method outlined in \cite{Mitarai2018}, which requires two extra repetitions of the circuit for each $\theta_i$:
\begin{equation}
\label{eqn:deriv}
\frac{\partial\langle C \rangle}{\partial \theta_i} = \frac{1}{2} \left( \langle C \rangle^+ - \langle C \rangle^- \right) ,
\end{equation}
where $\langle C \rangle^\pm$ is calculated by changing $\theta_ i$ by $\pm \frac{\pi}{2}$, and leaving all other parameters in the circuit constant.

\subsection{Reduced circuit}\label{sec:circuitdepth}
For the probability distribution of $a$ determined by $\mu_a = 0.25$ and $\sigma_a = 0.01$ the maximum theoretical success rate ($\textrm{P}_{\textrm{suc}}$) is $0.8333$ \cite{Mohseni2004}, which was obtained with the long circuit in Ref. \cite{Chen2018}.
However, after optimisation of circuit parameters for our larger circuit in Figure~\ref{fig:U_V_long} we reach only $0.72$, which is significantly smaller than the theoretical limit.
We attribute this discrepancy to the different implementations of the optimisation procedure, and to the different calculation of the gradients. 
To overcome this sub-optimal result we designed the shorter circuits in Fig. \ref{fig:U_V_short}. 
The choice of the reduced circuit is motivated by the consideration that for this task the rotations on the state qubits have a smaller effect on the measurement outcomes than rotations on the measurement qubits.
This choice of structure is so that the input states are entangled with both output qubits, and then the measurement qubits are rotated.
The choice of rotations about the x-axis, followed by the z-axis, and then again the x-axis allows for the initial state to be transformed to any other state on the surface of the Bloch sphere \cite{Nielsen2000a}.
With this short circuit (Figure~\ref{fig:U_V_short}) we obtain a success rate of $0.826$, close to optimal performance. 
This is the circuit used for the results presented, except where we explicitly note that the longer circuit is used.
\begin{figure*}
	\includegraphics[width=0.99\textwidth]{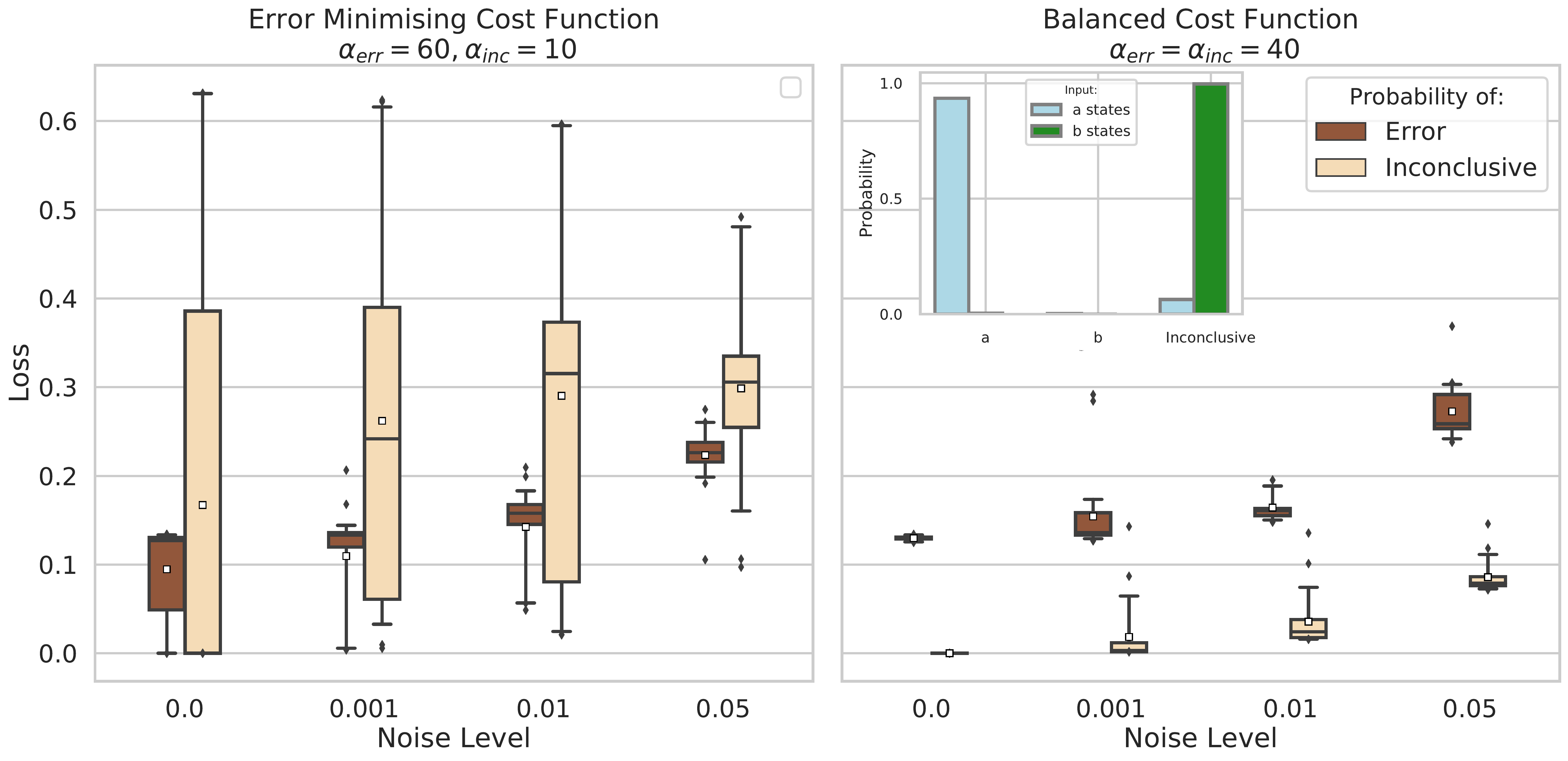}
	\caption{\label{fig:min_err} The distribution of $\textrm{P}_{\textrm{inc}}$ and $\textrm{P}_{\textrm{err}}$ from 25 repeats of a network biased towards reducing errors, and one with a balanced cost function, for $\mu_a=0.5$, and $\sigma_a=0.15$. An example undesirable output for a single minimising error run is in the inset, where no $b$ states are measured correctly, but the network still converges (the x-axis shows the output label and the colour is the input state). The interquartile range is contained within the box, and the 5th and 95th percentiles are marked by the whiskers. Outliers of this range are marked by a diamond. The mean is marked with a white square, and the median is the line across the box.}
\end{figure*}

We note that as the shorter circuits do not explore the full Hilbert space of all the qubits, they may not be necessarily optimal for all discrimination tasks.
Investigations into the capability of different variational quantum circuits have been made in \cite{Sim2019}.
Here we present evidence that when used on a noisy device, the smaller variational circuit converges to better results than the larger circuit. In general a trade-off needs to be made between this better resilience to noise and the ability of the circuit to distinguish very complex states.

\subsection{Noise}\label{sec:methods_noise}
Noise in quantum computers can be modelled by a superoperator, $\mathcal{E}(\rho)$, which is a completely positive, trace-preserving map on the state $\rho$ \cite{Nielsen2000a}.
We can give the operator-sum representation of $\mathcal{E}$ by introducing the Kraus operators, $E_k$:
\begin{equation}
\mathcal{E}(\rho) = \sum_k E_k\rho E_k^\dagger,
\label{eq:kraus1}
\end{equation}
and to preserve the trace of $\rho$, they must obey the relation
\begin{equation}
\sum_k E_k^\dagger E_k = \mathds{1}.
\end{equation}
For the single-qubit noise channel our operators are the single qubit Pauli operators, modified by the noise probability, $p$, to give the depolarising channel:
\begin{align}
\begin{split}
		E_0 &= \sqrt{1 - \frac{3p}{4}}\begin{bmatrix}
		1 & 0 \\
		0 & 1
	\end{bmatrix},\;
		E_1 = \sqrt{\frac{p}{4}}\begin{bmatrix}
		0 & 1 \\
		1 & 0
	\end{bmatrix}, \\
		E_2 &= \sqrt{\frac{p}{4}}\begin{bmatrix}
		0 & -i \\
		i & 0
	\end{bmatrix},\;\;\;\;\;\;
		E_3 = \sqrt{\frac{p}{4}}\begin{bmatrix}
		1 & 0 \\
		0 & -1
	\end{bmatrix}.
\end{split}
\label{eq:kraus2}
\end{align}

For the two qubit noise channel, which is applied after a two-qubit gate, the Kraus operators are tensor products of the combinations of these operators, i.e. $E_0 \otimes E_0, \dots E_1 \otimes E_2, \dots E_3 \otimes E_3$.	
The probability of the single qubit noise channel is $p_{1q} = \frac{4}{5}p_{2q}$.
This is the one-qubit marginal probability of error for the two-qubit gates \cite{Knill2004}, i.e. the probability of a single qubit error without condition of an error on the other qubit.
This is a commonly used assumption in the quantum error correction literature \cite{Gottesman2010}, which assumes that the error process  in single and two qubit gates is the same.
In real devices the process can be quite different, but we nevertheless choose this method as it is an upper limit on the error probability of the single qubit gate.
When quoting the noise level in this paper, we will always refer to $p_{2q}$. We set the highest noise level in our simulations to $p_{2q} = 0.1$, as this is an upper limit on two-qubit gate fidelities reported on current quantum hardware \cite{IBM2019, Hong2019, Chen2014}.

Note that here we have not considered asymmetric noise or different quality qubits.
However, we believe that correcting for a systematic bias such as this is possible for a variational algorithm, as seen in \cite{Mcclean2016}.
Furthermore, in actual devices the single qubit noise probability reported is much lower than $4/5$ of the two qubit gate noise level.
For example, the single qubit gate error rate reported in \cite{Chen2014} is $\num{1.4e-3}$, whereas the two qubit gate fidelity is $\num{9.3e-3}$, and the ratio between these is approximately $3/20$, at least a factor of $5$ lower.
In our simulations the single qubit noise is set to the higher limit of 4/5, so that we are more demanding of the algorithm.

\subsection{Simulation}
Simulations of the quantum device were performed on a simulator built using the Tensorflow machine learning package \cite{Abadi2016a}, and verified with the Cirq \cite{Gidney2018} quantum simulation package.
In our simulations we set the initial angles, which are our parameters to be optimised, at random values.
The labelled quantum state is an input to the circuit in Figure~\ref{fig:gen_circuit}, that circuit is ran and the measurement probabilities calculated and with them the cost.
The gradient of the cost with respect to each parameter is then calculated by the method described in section \ref{sec:methods_gradients}, and the parameters are updated according to the Adam optimiser to minimise the cost.
This routine is repeated until the cost no longer significantly decreases.

Measurements here are calculated in the `infinite-shot' regime, where the representation of the quantum state at the end of the circuit is used to extract exact measurement probabilities.
The inclusion of statistical measurement noise can be expected to result in slower rate of convergence than obtained here. We note that in Ref.  \cite{Sweke2019} it was demonstrated that convergence of variational algorithms is guaranteed even for single-shot measurements of the gradient. We indicate that convergence can also be achieved using this method in presence of measurement noise, although with a higher number of iterations.

\section{Results}\label{sec:results}
\subsection{Effect of cost function choice and circuit depth}\label{sec:wo_noise}
In Figure~\ref{fig:min_err} we compare the obtained optimised $\textrm{P}_{\textrm{err}}$ and $\textrm{P}_{\textrm{inc}}$ for an error minimising cost function ($\alpha_{err}=60,\alpha_{inc}=10$) and a balanced cost function ($\alpha_{err}=40,\alpha_{inc}=40$).
The error minimising cost function often results in a practically unusable network, because while it gives a low probability of error, the probability of inconclusive results is too high, as seen for an extreme case in the inset of Figure~\ref{fig:min_err}.
Note that in this particular case all $b$ states are detected as inconclusive, and one could in principle switch the \textit{inconclusive} and $b$ labels to obtain a good discrimination.
However, for the more general case this will not be possible.

In comparison to the error minimising setting, the results for the balanced cost function are stable and generally give both small $\textrm{P}_{\textrm{err}}$ and $\textrm{P}_{\textrm{inc}}$, with some $\textrm{P}_{\textrm{err}}$ comparable to the error minimising setting.
For the remaining analysis we therefore use the balanced cost function ($\alpha_{\textrm{err}} = \alpha_{\textrm{incon}} = 40$). 
We note that as the noise level is increased, $\textrm{P}_{\textrm{inc}}$ and $\textrm{P}_{\textrm{err}}$ progressively tend to larger values.
The effect of noise will be analysed in detail in the next section. 

\begin{figure}[!bht]
	\centering
	\includegraphics[width=0.45\textwidth]{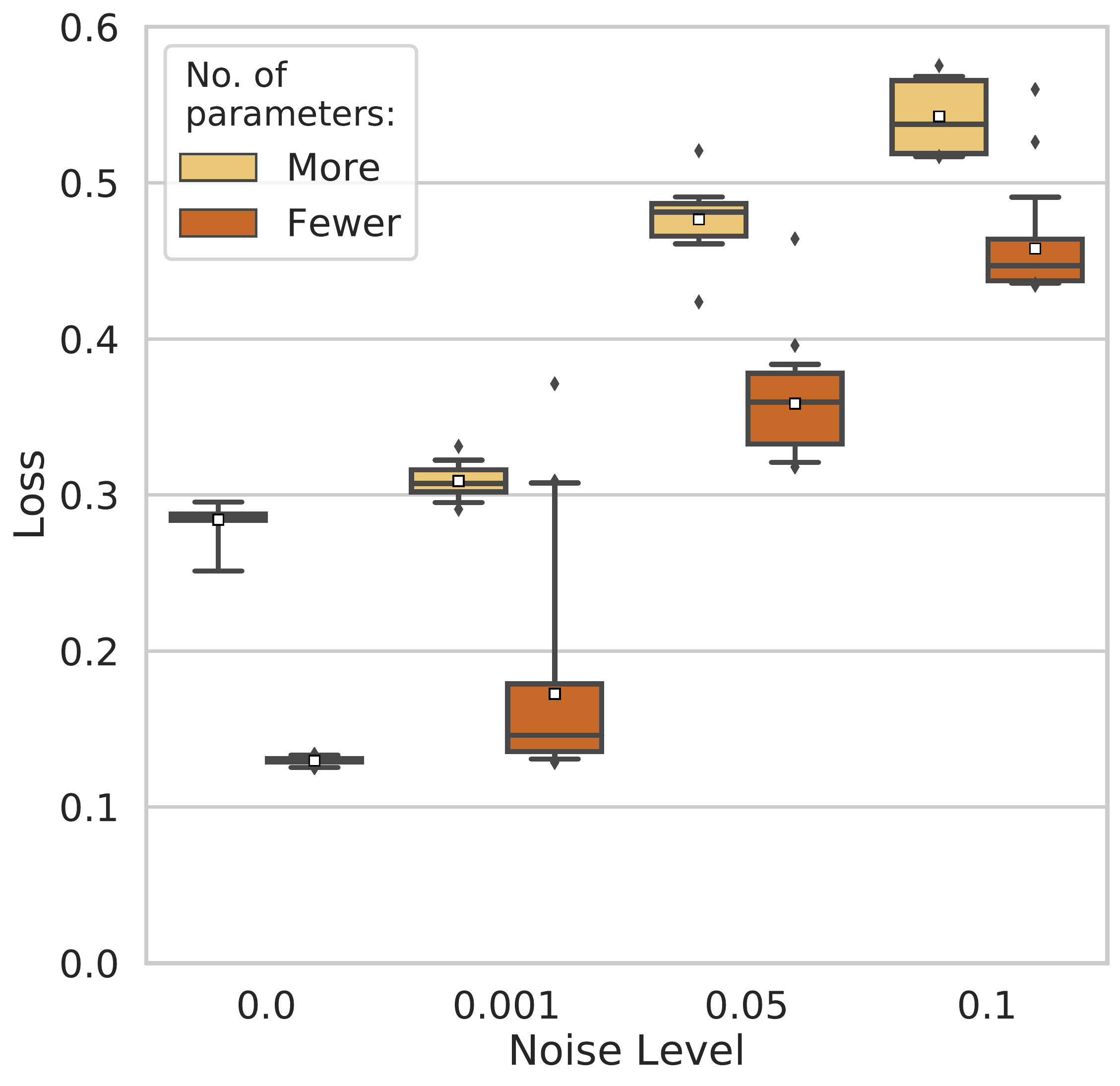}
	\caption{\label{fig:circ_len_compare} The distributions of loss ($\textrm{P}_{\textrm{err}} + \textrm{P}_{\textrm{inc}}$)  at different noise levels for the two circuits shown in Figure~\ref{fig:U_V_circuits}. Both have other parameters fixed, $\mu_a = 0.5, \sigma_a = 0.15, \alpha_{\textrm{err}} = \alpha_{\textrm{inc}} = 40$. We observe that reducing the number of parameters is advantageous at all noise levels.}
\end{figure}
We next investigate the influence of the number of parameters in the quantum circuit on the loss. In Figure~\ref{fig:circ_len_compare} we compare the distributions of loss between the circuit with more trainable parameters in Figure~\ref{fig:U_V_long} to the circuit with fewer parameters in Figure~\ref{fig:U_V_short}. It can be seen that the reduced circuits consistently perform better than the long circuits.
\begin{figure}[!htb]
	\centering
	\includegraphics[width=0.45\textwidth]{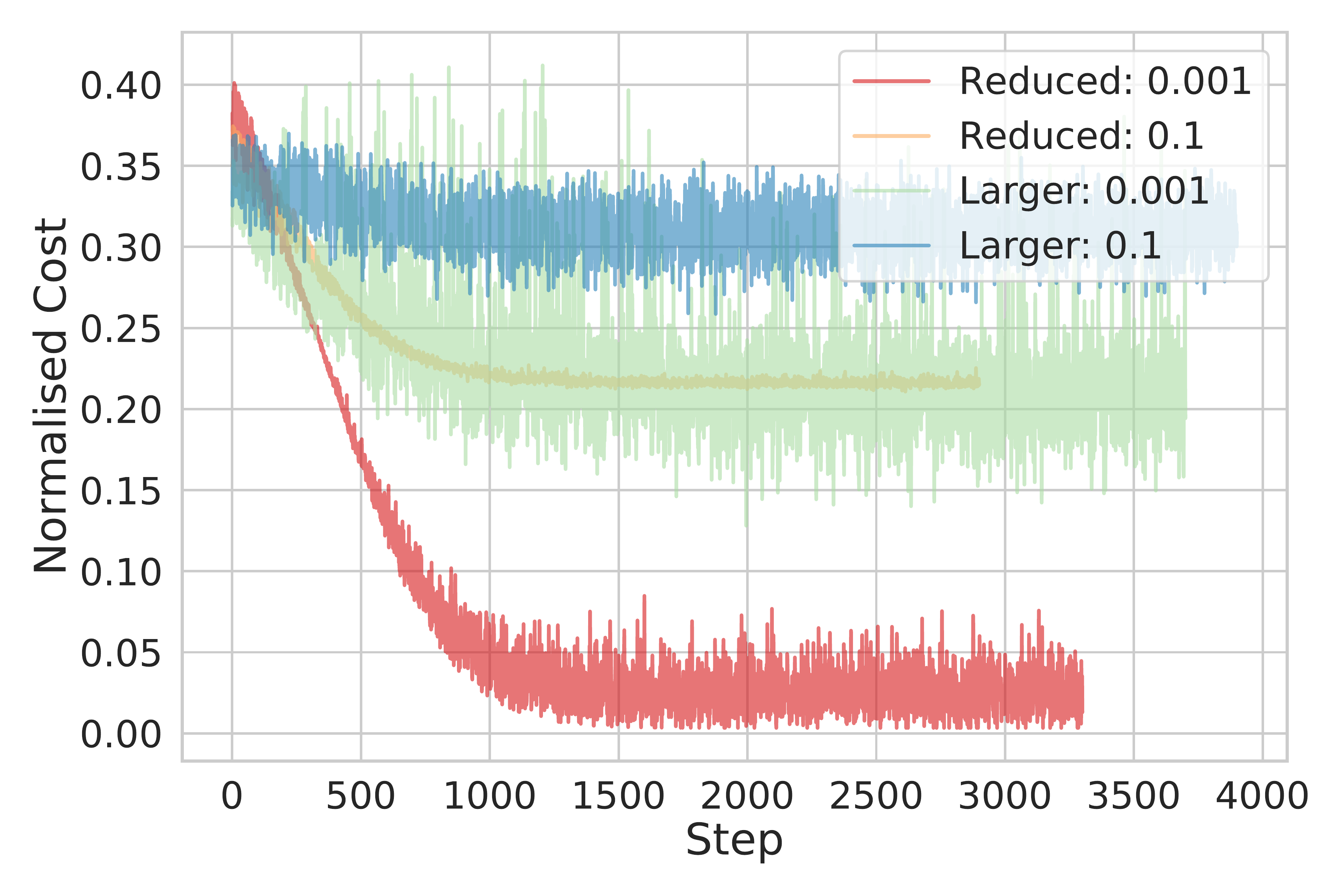}
	\caption{\label{fig:loss_fns_all} Evolution of the normalised cost functions for larger and reduced circuits for $\mu_a = 0.5$ and $\sigma_a = 0.15$, with noise levels of $0.001$ and $0.1$. Shown here is the number of steps taken to converge. Note that the time taken to complete a single step of the longer circuit is much greater than for the reduced circuit.}
\end{figure}
It is more difficult to train circuits with a large number of parameters both without and with noise, as seen in Figure~\ref{fig:loss_fns_all}.
We see that the higher noise cases always converge to a higher loss, and that the reduced circuits perform better in both cases.
These results show that in practice increasing the number of parameters used in a quantum circuit does not always have a beneficial effect. Importantly, even in the noiseless case the circuit with less parameters leads to better results. Furthermore the reduced number of parameters also significantly lowers the required run-time.

Even with very low noise, the output is worse for larger circuits.
This suggests that with more parameters the algorithm struggles to optimise, when the gradient calculations are performed on the quantum device.
Good performance of the short circuit in the presence of noise can be due to the noisy gradient regularising the training, thereby optimising performance \cite{Goodfellow2016}.
Moreover, the Adam optimiser has been designed to work well with noisy gradients \cite{Kingma2014}.
The results seen here are indicative that a noise-resilient optimiser using gradients provided by a noisy quantum circuit can perform well.

\subsection{Effect of noise: numerical analysis and model}\label{sec:w_noise}

\begin{figure}[!htb]
	\centering
	\includegraphics[width=0.49\textwidth]{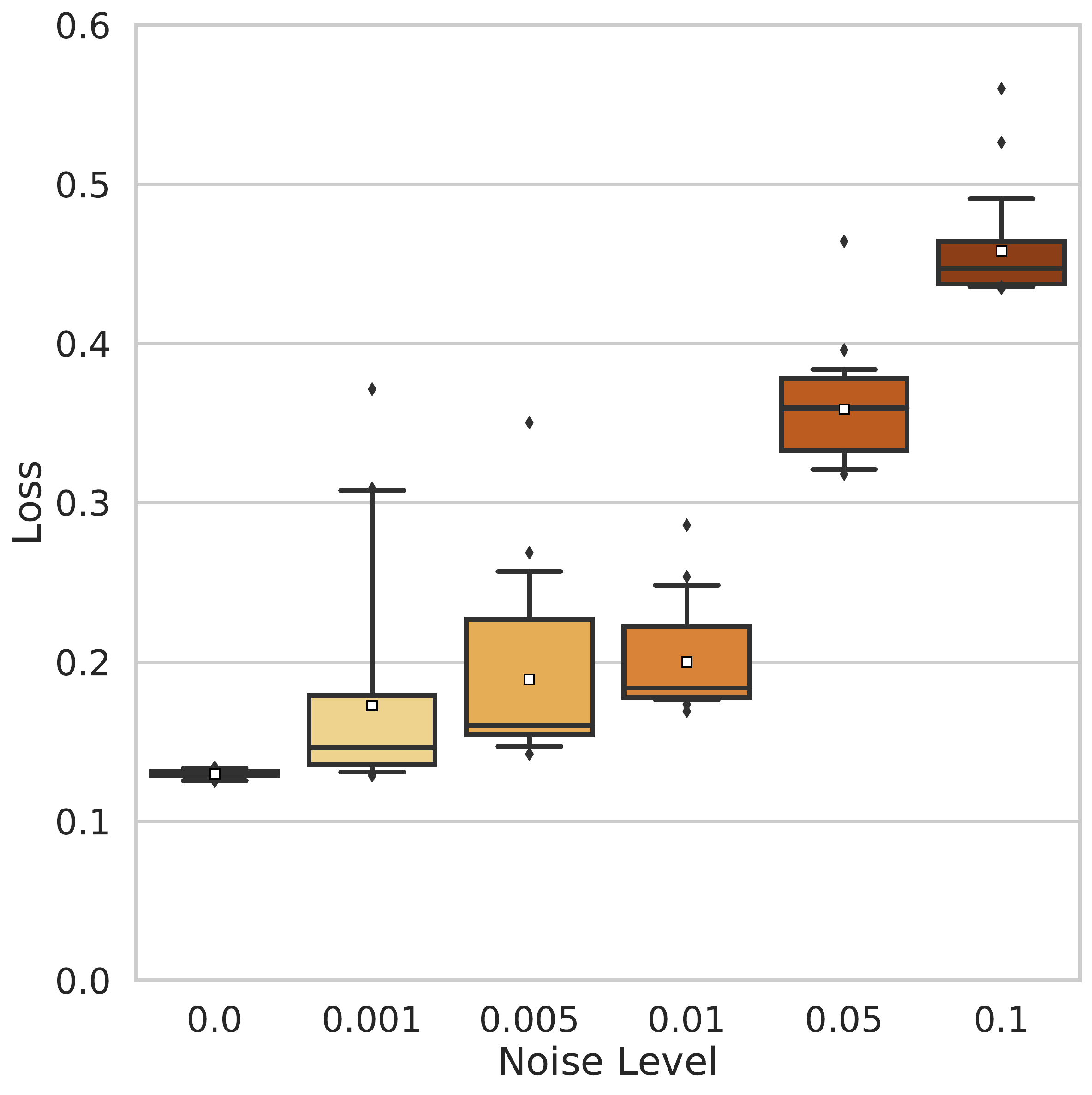}
	\caption{\label{fig:noise_array} The distribution of loss for 25 repeats of training the network. The cost function is balanced, $\alpha_{\textrm{err}}=\alpha_{\textrm{inc}}=40$, $\mu_a=0.5$, and $\sigma_a=0.15$. At levels of noise present in current devices, $0.01$, the loss value is favourable, an average of $0.2$.}
\end{figure}
In Figure~\ref{fig:noise_array} the noiseless case is compared to resulting optimised loss for increasing noise levels (note that in this section we always use the reduced circuit). It can be seen that using this algorithm with zero noise produces the lowest loss, as one expects intuitively. With increasing noise the average loss increases continuously. In presence of noise there are a few high-loss outliers, which we attribute to the optimiser becoming stuck in local minima of the cost function.
As the noise is increased, performance deteriorates, but is no worse than the random output limit of $2/3\approx0.67$ (see Sec. \ref{sec:methods_QNN}).

Importantly, at noise levels comparable to current devices, $p_{2q} = 0.01$, the algorithm is still performing well, at an average loss of $0.2$.
\begin{figure}[!tb]
	\centering
	\includegraphics[width=0.49\textwidth]{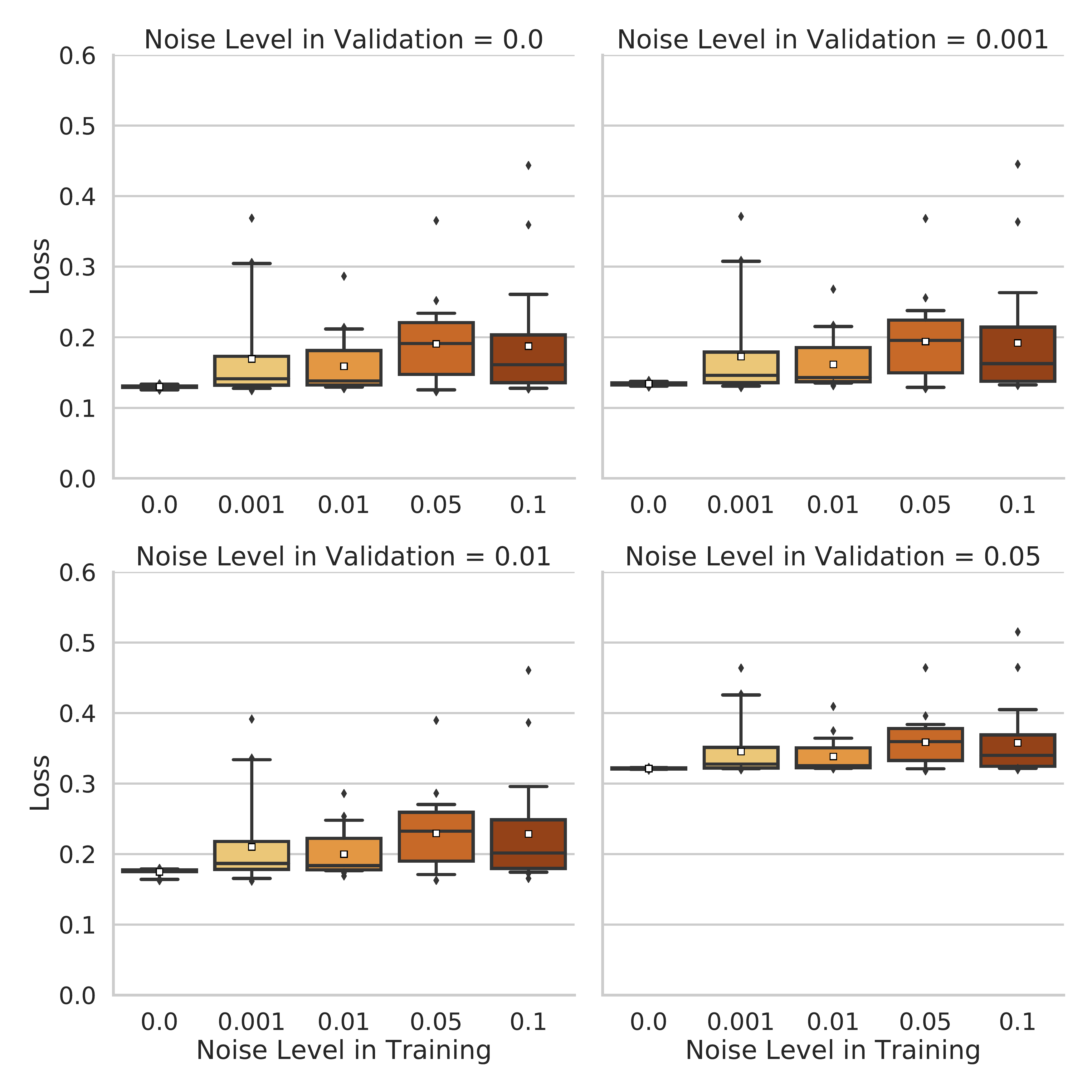}
	\caption{\label{fig:all_noise_plot} Distribution of loss ($\textrm{P}_{\textrm{err}} + \textrm{P}_{\textrm{inc}}$) against training noise for a selection of noise levels in the validation circuit.}
\end{figure}

\begin{figure}[!htb]
	\centering
	\includegraphics[width=0.49\textwidth]{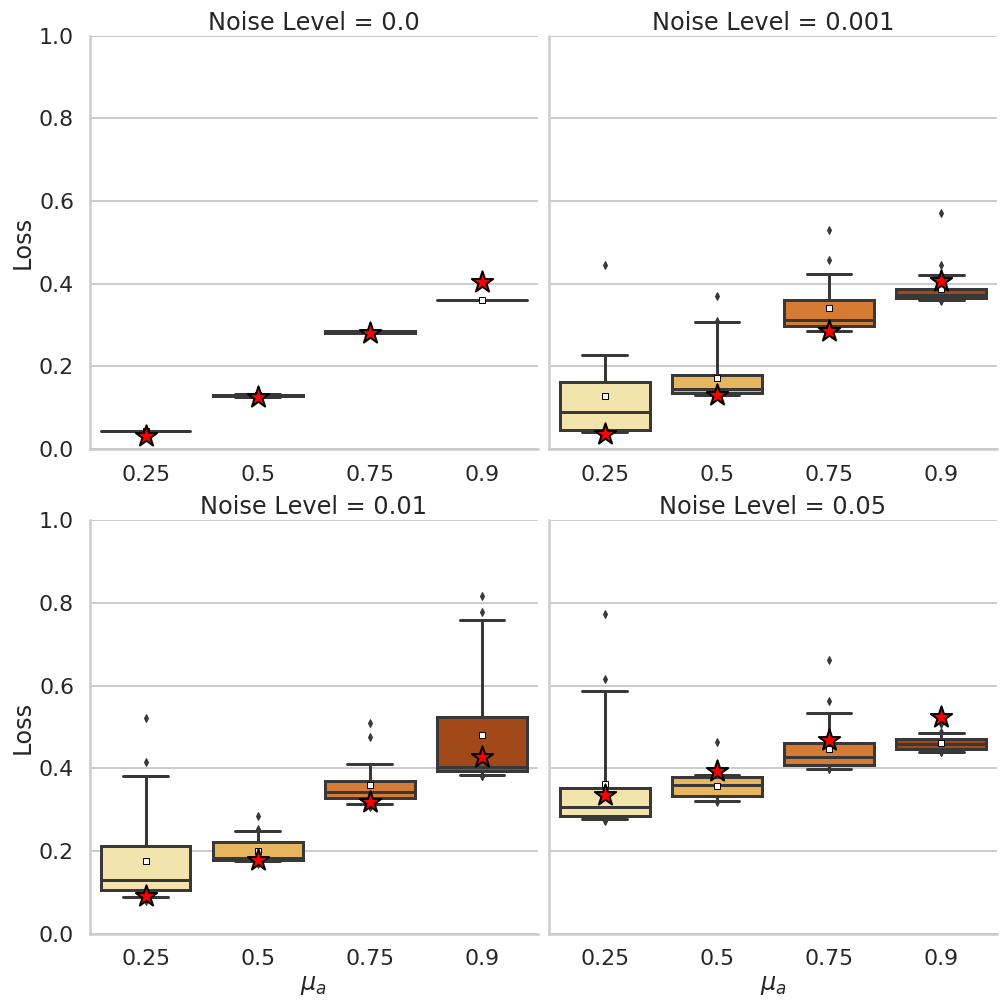}
	\caption{\label{fig:mu_increasing} The distribution of loss ($\textrm{P}_{\textrm{err}} + \textrm{P}_{\textrm{inc}}$) and the effect of different values of $\mu_a$. The cost function is balanced, $\alpha_{\textrm{err}}=\alpha_{\textrm{inc}}=40$, and $\sigma_a=0.15$. We see that for lower values of $\mu_a$, corresponding to smaller overlap between the states to be discriminated, the discrimination task is performed better. The red stars indicate the fidelity $F_{\tilde{a}\tilde{b}}$ between the two states after three applications of Kraus operators to each of the data qubits, as given by Eq.~(\ref{eqn:fidelity_both}).}
\end{figure}

In general a high level of noise always leads to a higher loss. However, we find that when noise is applied only during the training of the parameters, the optimised parameters are rather resilient to this training noise. To show this in
Figure~\ref{fig:all_noise_plot} we present the results when training the device at one noise level, and validating at another. We see that even with high levels of training noise the optimiser converges onto good parameters, as we find comparably low loss levels when validating those parameters trained at a high noise level with low noise in the validation step.
Also here we find that when validating at noise levels seen in current devices, $p_{2q} = 0.01$, the average loss does not increase above $0.25$, which would be acceptable to use for state discrimination.

In order to provide an understanding of the numerically found changes of the loss with noise, in what follows we present a simple model that can describe the results. It is based on the notion that a larger overlap between the states to be discriminated between generally makes discrimination more difficult. As outlined in Sec. \ref{sec:methods_qsd}, with our chosen set of states this overlap can be tuned by setting the value of $a$, and is equal to $a/\sqrt{2}$. We can therefore systematically evaluate the effect of noise on the discrimination for increasing overlap by increasing $\mu_a$, and the results are shown in Figure~\ref{fig:mu_increasing}. The loss increases for larger $\mu_a$ for all levels of noise.
At high noise levels and high $\mu_a$, some runs are performing even worse than the random output limit ($0.67$), but on average the loss remains well below that value.
In general we conclude that the tolerable levels of noise depend on the overlap between the states, where small overlap allows the states to be discriminated even for higher noise in the quantum computer.

\begin{figure*}[ht]
    \begin{subfigure}[b]{0.29\textwidth}
        \caption{\label{fig:fidelity_aa}}
        \includegraphics[width=\textwidth]{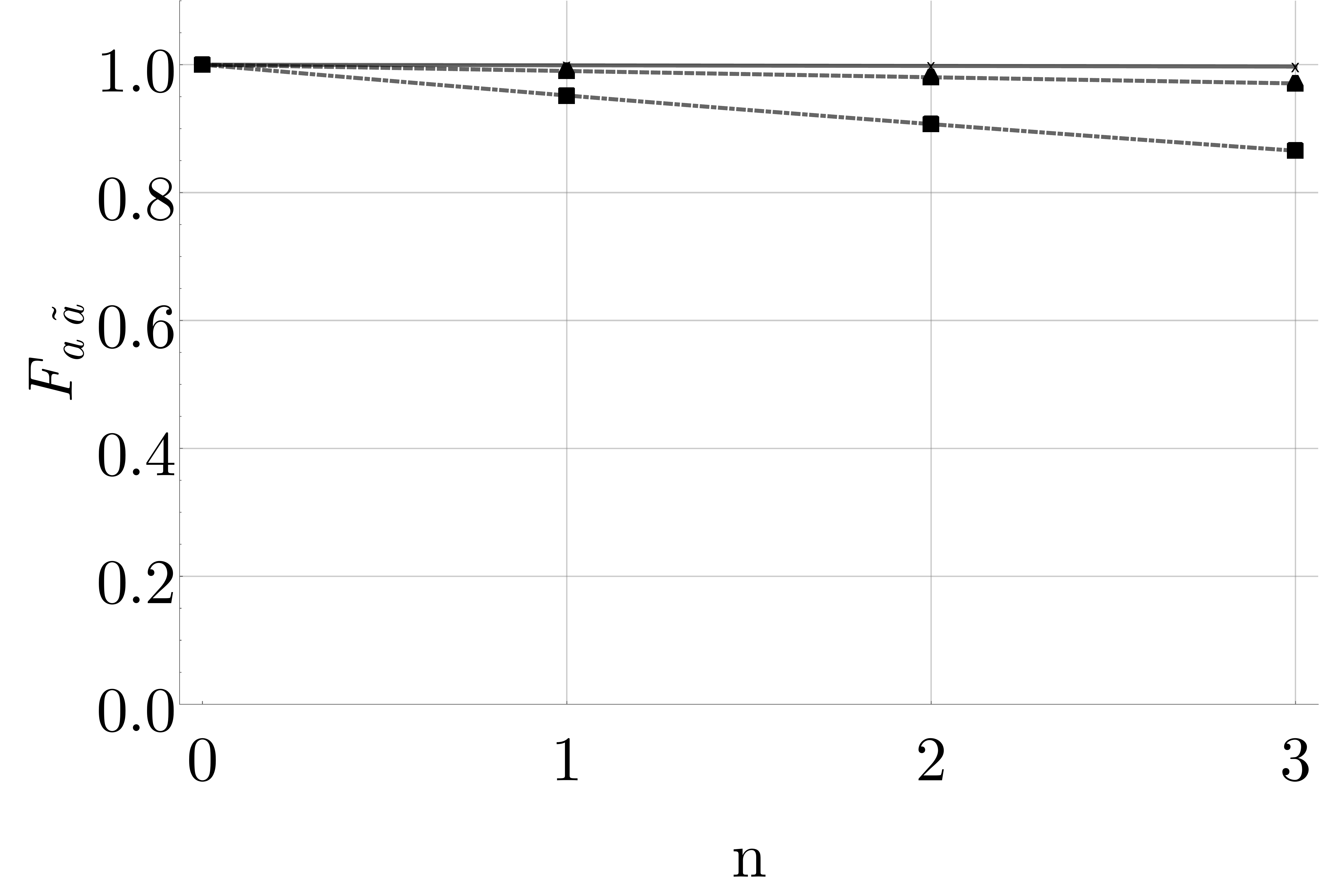}
    \end{subfigure}
    \begin{subfigure}[b]{0.29\textwidth}
        \caption{\label{fig:fidelity_noisyab}}
        \includegraphics[width=\textwidth]{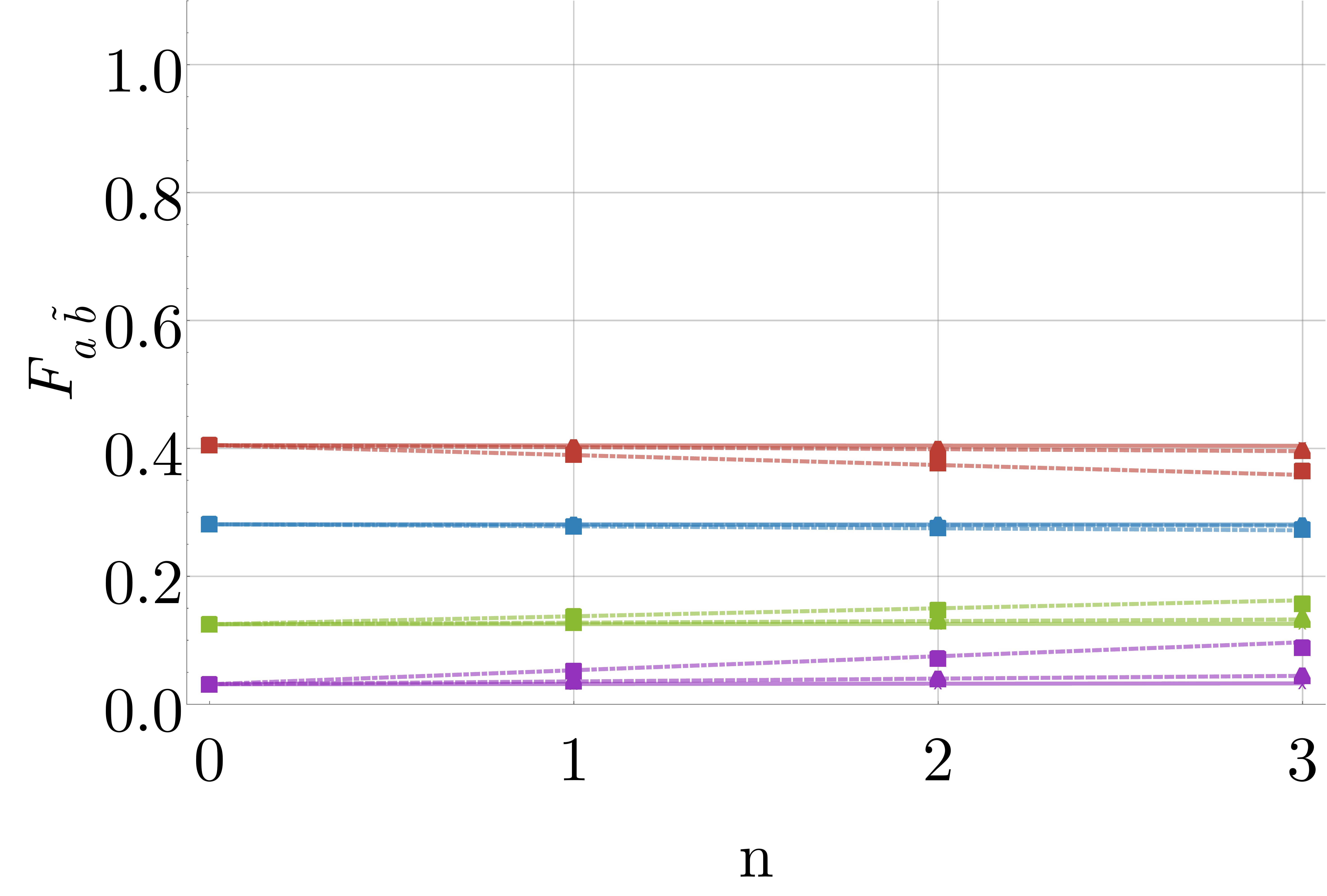}
    \end{subfigure}
    \begin{subfigure}[b]{0.29\textwidth}
        \caption{\label{fig:fidelity_both}}
        \includegraphics[width=\textwidth]{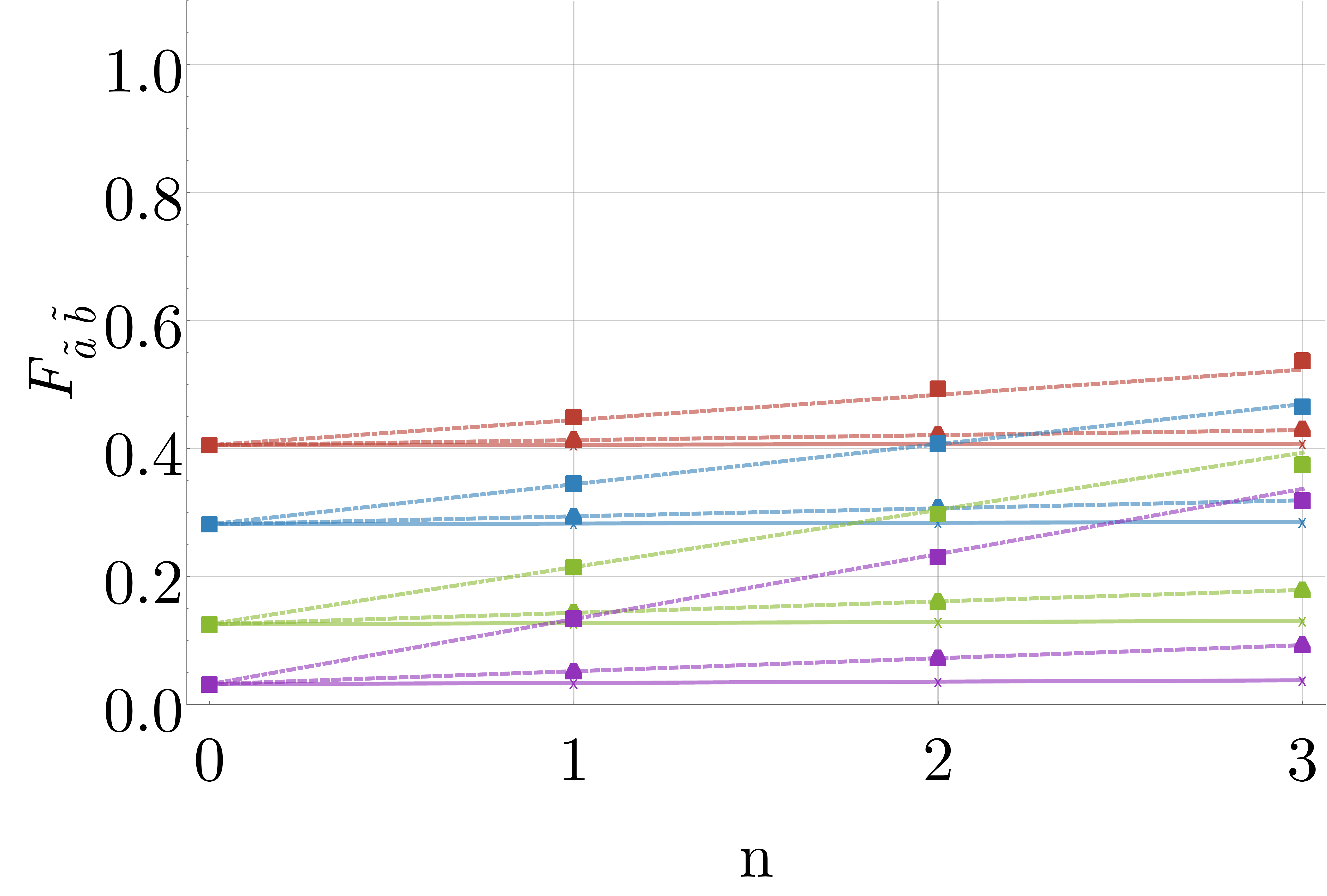}
    \end{subfigure}
    \begin{subfigure}[b]{0.1\textwidth}
        \includegraphics[width=\textwidth]{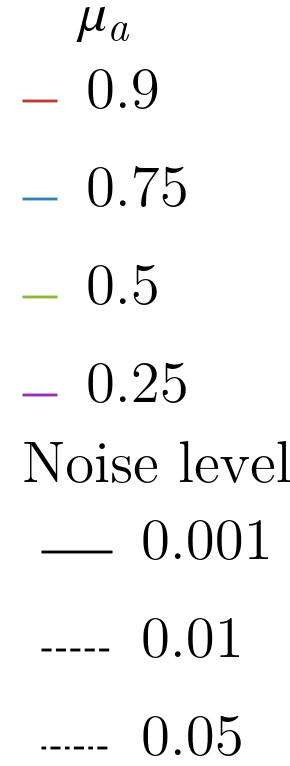}
    \end{subfigure}
	\caption{\label{fig:fidelity}Fidelities as function of the number of applied noise channels, $n$, between (a) the same states with noise applied to one state ($F_{a\tilde{a}}$), (b) the two different states with noise applied to one state ($F_{a\tilde{b}}=F_{b\tilde{a}}$), and (c) with noise applied to both states ($F_{\tilde{a}\tilde{b}}$). Markers show the calculated numeric fidelities using Eq. (\ref{eqn:fidelity}), lines show the low order expansions given by  Eq.~(\ref{eqn:fidelity_fit}). The low order expansion agrees well with the numerical results for all cases.}
\end{figure*}

For large noise in the system the difference in loss between higher and lower values of $\mu_a$ is significantly reduced when compared to the low noise case. 
This seems to indicate that the noise on average reduces the difference between states as these pass the circuit, and hence effectively increases the averaged overlap.
This effect can be illustrated for the ideal case of $a=0$ and no noise, where discrimination can in principle be perfect since the states are orthogonal.
However, in presence of noise there is a probability that a state is perturbed as the circuit is applied to it, and hence orthogonality between states is lost.
This results in a certain probability of erroneous detection.

In order to estimate this effect on a semi-quantitative level for our used circuit, shown in Figures~\ref{fig:gen_circuit} and~\ref{fig:U_V_short}, we note that in absence of noise the role of the data qubits is only to store the state $\left.|\psi_{\textrm{in}}\right>$, which then controls the state of the measurement qubits. In presence of noise the three noisy two-qubit gates applied to each of the two data qubits will perturb $\left.|\psi_{\textrm{in}}\right>$ during the processing of the circuit, which in turn will affect the measurement qubits via the control operation and hence the outcome of the state discrimination. We therefore approximately model the effect of noise on the state discrimination by determining how much the repeated application of noisy channels at each gate affects a given $\left.|\psi_{\textrm{in}}\right>$, without considering the presence of the measurement qubits.

In what follows we quantify how the application of noise channels affects each state and its overlap with the state to be discriminated from.
The quantum states with noise applied are represented by density matrices, so that the overlap between two states represented by the density matrices $\rho$ and $\sigma$, respectively, is described by the fidelity, $F$, given by \cite{Jozsa94}:
\begin{equation}
    \label{eqn:fidelity}
    F =  \textrm{Tr}\left[\sqrt{\sqrt{\sigma} \rho \sqrt{\sigma}}\right]^2.
\end{equation}
Of particular interest here is the fidelity between a pure state $\ket{\psi_a}$ entering the circuit and its modified form due to the application of noise after three two-qubit gates.
We denote this as $F_{a\tilde{a}}$, where the tilde on the second subscript indicates that the second state is the one where noise was applied. 
We use an analogous notation the other relevant quantities, which are $F_{b\tilde{b}}$, $F_{b\tilde{a}}$ and $F_{\tilde{b}\tilde{a}}$.
We can compute these quantities numerically, but given the rather cumbersome form of Eqn.~(\ref{eqn:fidelity}) it is difficult to relate the results to the fundamental parameters of the noisy discrimination process.
We therefore provide a lowest order expansion of these terms in $p$, which we expect to be close to the exact results since we are only dealing with small $p$.
We apply the noise model in Eqs.~(\ref{eq:kraus1}-\ref{eq:kraus2}) to the states $a$ and $b$ in Eqs.~(\ref{eqn:states_a}) and (\ref{eqn:states_b}), and obtain to first order in $p$.
\begin{subequations}
    \label{eqn:fidelity_fit}
    \begin{equation}
        \label{eqn:fidelity_a_on_a}
        F_{a\tilde{a}} = 1 - n\; p\;,
    \end{equation}
    \begin{equation}
        \label{eqn:fidelity_b_on_b}
        F_{b\tilde{b}} = 1 - \frac{3n\; p\;}{2},
    \end{equation}
    \begin{equation}
        \label{eqn:fidelity_a_on_b}
        F_{a\tilde{b}} = F_{b\tilde{a}} = \frac{\mu_a^2}{2} + \frac{n\; p}{2}\left(1 - 2\mu_a^2 \right),
    \end{equation}
    \begin{equation}
        \label{eqn:fidelity_both}
        F_{\tilde{a}\tilde{b}}  = \frac{\mu_a^2}{2} + n\; p\;\left(1 + \frac{\mu_a}{\sqrt{2}} - 2\mu_a^2 + \sqrt{1 - \frac{\mu_a^2}{2}} \right),
    \end{equation}
\end{subequations}
where $n = 0 \dots 3$ is the number of noisy channel applications, $\mu_a$ is the mean value of $a$ and $p$ is the noise probability. 
Within this expansion order the fidelities are linear in both $p$ and $n$, and the expansion coefficients are simple functions of $\mu_a$. 
For the noiseless case ($p=0$), $F_{a\tilde{b}}=F_{\tilde{a}\tilde{b}}=\mu_a^2/2$, which corresponds to the absolute value squared of the overlap between the $a$ and $b$ states.

The results for the fidelities for increasing $n$ are shown in Figure \ref{fig:fidelity}. It can be seen that the numerical results obtained directly with Eqn.~(\ref{eqn:fidelity}) are captured rather well with the the analytical low order expansion in Eqs.~(\ref{eqn:fidelity_a_on_a}-\ref{eqn:fidelity_both}). 
The value of $F_{a\tilde{a}}$ decreases with each application of a noisy channel, showing that the purity of the state degrades as the noise channels act on the state.
The value of $F_{a\tilde{b}}$ increases with $n$ for lower values of $\mu_a$, where the noise acts to increase the fidelity between the states, while it decreases with $n$ for higher values of $\mu_a$, where the noise reduces the orthogonal component in $\ket{\psi_a}$.
In Figure~\ref{fig:fidelity_both} the fidelity is plotted when the noise channel is applied to both states ($F_{\tilde{a}\tilde{b}}$), and it always increases with $n$.
$F_{\tilde{a}\tilde{b}}$ is the relevant quantity the influence of noise on the loss: the application of the noisy circuit on the originally pure input states causes them to degrade into mixed states. 
The states to be discriminated are therefore not the input pure states anymore, but these noisy states, and for each application of the noise channel they become harder to discriminate.

The minimal loss achievable in quantum state discrimination is generally a function of the fidelity between the two input states \cite{Barnett09, Bae17}. The exact relation depends on the cost function that is minimised, and is only known analytically for a few special cases, such as minimal error discrimination or unambiguous discrimination \cite{Barnett09, Bae17}. In this section we consider the case, where the rate of inconclusive results and erroneous results is minimised simultaneously ($\alpha_{\textrm{err}} = \alpha_{\textrm{inc}} = 40$). Before estimating the effect of noise we therefore need to determine the functional relation between the fidelity and the loss in the noiseless case for our circuit. In the top left panel of Fig. \ref{fig:mu_increasing} we show the results for $F_{\tilde{a}\tilde{b}}(p=0)=F_{ab}$ as red stars for each $\mu_a$. One can see that the $F_{ab}$ is approximately equal to the loss for all $\mu_a$, so that to a good approximation for our circuit we can fit the relation as $\mathrm{Loss}=F_{ab}$.
In general the lower bound of the optimal theoretical loss is found in the minimum error discrimination setting, where there is no inconclusive measurement.
In our optimisation we include also the inconclusive measurement, the probability of which is minimised in the unambiguous setting, which gives an upper bound on the loss.
Our minimised cost function is a combination of these settings, which simultaneously minimises errors and inconclusive results, and our relation for the loss in fact lies in between the optimum values in each of these boundary settings.

We can now verify the validity of our model for the effect of noise on the loss. To this aim we calculate $F_{\tilde{a}\tilde{b}}$ from Eqn.~(\ref{eqn:fidelity_both}) for $n=3$, which corresponds to the number of entangling gates applied in our circuit to each data qubit.
These results are presented in Figure~\ref{fig:mu_increasing} for the panels with $p\ne0$. 
One can see that $F_{\tilde{a}\tilde{b}}$ agrees rather well with the loss also in the presence of noise.
This validates our model that the effect of noise on the state discrimination is mainly determined by the noise-induced increased overlap between the states as they are processed in the circuit. 
In particular for the highest noise the model captures well the fact that the effect of noise is large for small $\mu_a$, while it is reduced for larger $\mu_a$. 
Eqn.~(\ref{eqn:fidelity_both}) therefore allows us to estimate the minimal loss achievable with our circuit for a given $p$ and $\mu_a$.

\begin{figure}[!htb]
	\centering
	\includegraphics[width=0.49\textwidth]{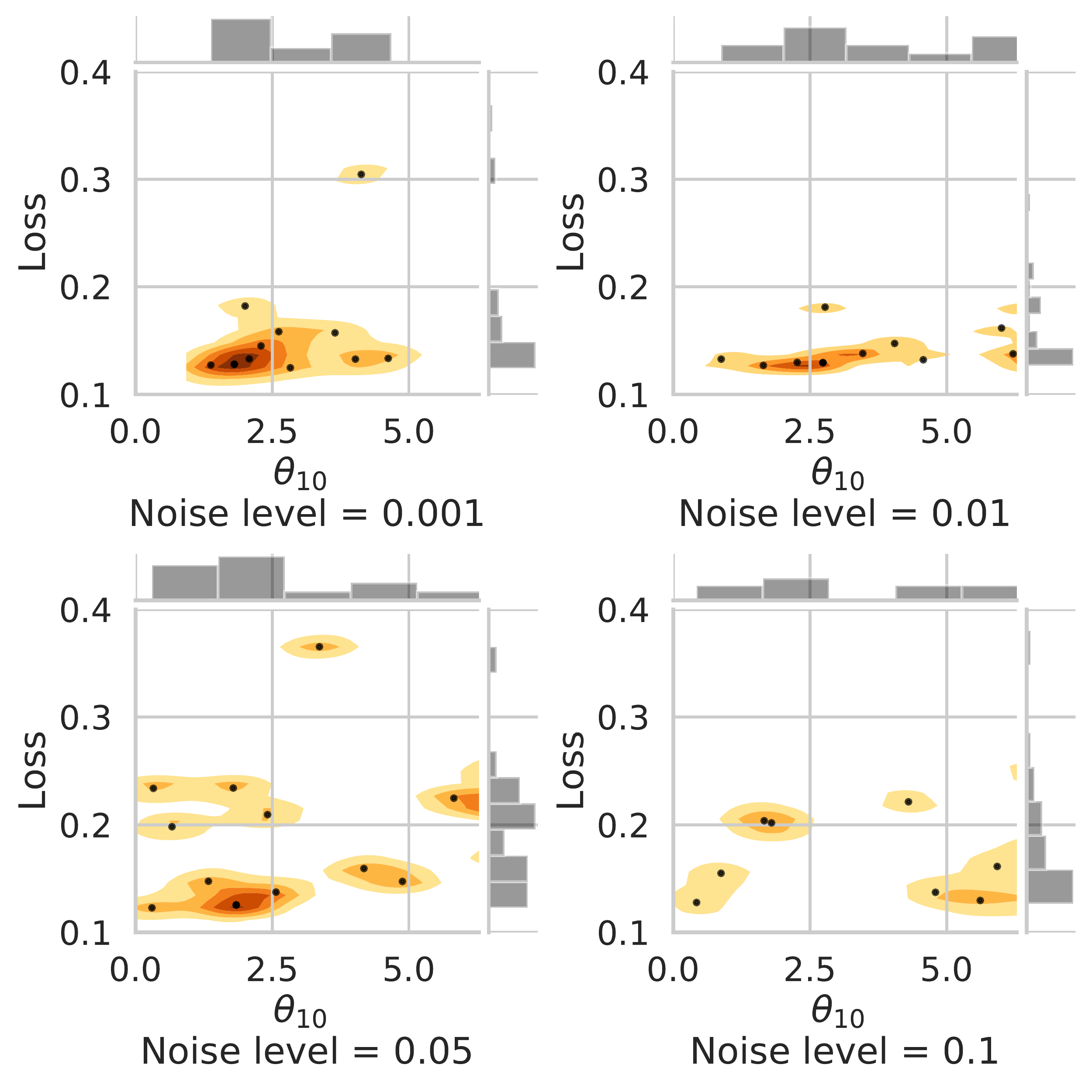}
	\caption{\label{fig:angles} The distribution of loss and $\theta_{10}$ obtained at different noise levels. Here we can see the effect of noise on the values of $\theta_{10}$ that the optimiser converges to. We only show a single representative parameter $\theta_{10}$, since we have a approximately similar behaviour for all other parameters.}
\end{figure}
Finally, we investigate the effect of the noise during training on the actual values of the optimised parameters in the circuit. In Figure~\ref{fig:angles} we present the distribution of $\theta_{10}$ for different values of noise.
The values of $\theta$ are all taken modulo $2\pi$, and at zero validation noise to remove the effect of validation errors on the loss.
We see that the range of angles converged upon increases as the noise in the circuit increases.
Some values become stuck at high loss, and there can be different values for the minimal loss parameters.
The increase in noise seems to change not just the final loss, but the parameters found that minimise loss.
We cannot rule out the correlation between different parameters as the noise level changes.
Combined with what we see in Figure~\ref{fig:all_noise_plot}, that good parameters are still found at higher noise levels, we may conclude that noise in the circuit can push the optimiser out of local minima, so that it can find some other local minima at lower loss.

From the results presented here we see that this algorithm performs well in the presence of noise in the training and validation steps (Figure~\ref{fig:noise_array}), and that parameters found on a noisy device work well when validated on a device with low noise (Figure~\ref{fig:all_noise_plot}).
When calculating parameter gradients on a noisy quantum device, reducing the number of parameters has a positive effect, as shown in Figure~\ref{fig:circ_len_compare} and Figure~\ref{fig:loss_fns_all}.

\section{Conclusion}\label{sec:conclusion}
We have shown that a QNN can be trained for the task of state discrimination on a noisy device, with noise levels found in current NISQ devices.
We have also shown that gradient descent algorithms are viable on noisy quantum devices, given a good choice of classical algorithm.
As discussed in \cite{Sim2019}, choice of training circuits in variational quantum algorithms has a large effect upon success. Here we reduced the number of parameters by removing rotation gates from the input states, and indeed show that the low circuit depth and qubit count of our algorithm is beneficial in the presence of noise.
We also developed a simple model equation relating the loss to the noise level and input state, which is based on the fact that in our circuit the application of noisy gates effectively leads to an increase of the overlap between states to be discriminated during the processing of the circuit.
While we specifically considered the task of quantum state discrimination, the algorithm presented here can be equally applied to such problems as verification of general quantum machine learning outputs, and applications in sensing, imaging and metrology.

\section*{Acknowledgements}
AP is supported by the InQUBATE Training and Skills Hub [grant EPSRC EP/P510270/1]. 
IR acknowledges financial support from the UK Department of Business, Energy and Industrial Strategy
(BEIS). 
HC acknowledges the support though a Teaching Fellowship from UCL.
LW acknowledges kindly the support through the Google PhD Fellowship in Quantum Computing.
DB acknowledges support from the EPSRC Prosperity Partnership in Quantum Software for Modelling and Simulation [grant No. EP/S005021/1].
AP acknowledges the use of the UCL Myriad High Throughput Computing Facility (Myriad@UCL), and associated support services, in the completion of this work.

\singlespacing
\raggedright 
\bibliography{abbreviated}

\begin{thebibliography}{38}%
\makeatletter
\providecommand \@ifxundefined [1]{%
 \@ifx{#1\undefined}
}%
\providecommand \@ifnum [1]{%
 \ifnum #1\expandafter \@firstoftwo
 \else \expandafter \@secondoftwo
 \fi
}%
\providecommand \@ifx [1]{%
 \ifx #1\expandafter \@firstoftwo
 \else \expandafter \@secondoftwo
 \fi
}%
\providecommand \natexlab [1]{#1}%
\providecommand \enquote  [1]{``#1''}%
\providecommand \bibnamefont  [1]{#1}%
\providecommand \bibfnamefont [1]{#1}%
\providecommand \citenamefont [1]{#1}%
\providecommand \href@noop [0]{\@secondoftwo}%
\providecommand \href [0]{\begingroup \@sanitize@url \@href}%
\providecommand \@href[1]{\@@startlink{#1}\@@href}%
\providecommand \@@href[1]{\endgroup#1\@@endlink}%
\providecommand \@sanitize@url [0]{\catcode `\\12\catcode `\$12\catcode
  `\&12\catcode `\#12\catcode `\^12\catcode `\_12\catcode `\%12\relax}%
\providecommand \@@startlink[1]{}%
\providecommand \@@endlink[0]{}%
\providecommand \url  [0]{\begingroup\@sanitize@url \@url }%
\providecommand \@url [1]{\endgroup\@href {#1}{\urlprefix }}%
\providecommand \urlprefix  [0]{URL }%
\providecommand \Eprint [0]{\href }%
\providecommand \doibase [0]{http://dx.doi.org/}%
\providecommand \selectlanguage [0]{\@gobble}%
\providecommand \bibinfo  [0]{\@secondoftwo}%
\providecommand \bibfield  [0]{\@secondoftwo}%
\providecommand \translation [1]{[#1]}%
\providecommand \BibitemOpen [0]{}%
\providecommand \bibitemStop [0]{}%
\providecommand \bibitemNoStop [0]{.\EOS\space}%
\providecommand \EOS [0]{\spacefactor3000\relax}%
\providecommand \BibitemShut  [1]{\csname bibitem#1\endcsname}%
\let\auto@bib@innerbib\@empty
\bibitem [{\citenamefont {Bennett}(1992)}]{Bennett1992}%
  \BibitemOpen
  \bibfield  {author} {\bibinfo {author} {\bibfnamefont {C.~H.}\ \bibnamefont
  {Bennett}},\ }\href {\doibase 10.1103/PhysRevLett.68.3121} {\bibfield
  {journal} {\bibinfo  {journal} {Phys. Rev. Lett.}\ }\textbf {\bibinfo
  {volume} {68}},\ \bibinfo {pages} {3121} (\bibinfo {year}
  {1992})}\BibitemShut {NoStop}%
\bibitem [{\citenamefont {Chefles}(2000)}]{Chefles2000}%
  \BibitemOpen
  \bibfield  {author} {\bibinfo {author} {\bibfnamefont {A.}~\bibnamefont
  {Chefles}},\ }\href {\doibase 10.1080/00107510010002599} {\bibfield
  {journal} {\bibinfo  {journal} {Contemp. Phys.}\ }\textbf {\bibinfo {volume}
  {41}},\ \bibinfo {pages} {401} (\bibinfo {year} {2000})}\BibitemShut
  {NoStop}%
\bibitem [{\citenamefont {Duan}\ and\ \citenamefont {Guo}(1998)}]{Duan1998}%
  \BibitemOpen
  \bibfield  {author} {\bibinfo {author} {\bibfnamefont {L.-M.}\ \bibnamefont
  {Duan}}\ and\ \bibinfo {author} {\bibfnamefont {G.-C.}\ \bibnamefont {Guo}},\
  }\href {\doibase 10.1103/PhysRevLett.80.4999} {\bibfield  {journal} {\bibinfo
   {journal} {Phys. Rev. Lett.}\ }\textbf {\bibinfo {volume} {80}},\ \bibinfo
  {pages} {4999} (\bibinfo {year} {1998})}\BibitemShut {NoStop}%
\bibitem [{\citenamefont {Giovannetti}\ \emph {et~al.}(2011)\citenamefont
  {Giovannetti}, \citenamefont {Lloyd},\ and\ \citenamefont
  {Maccone}}]{Giovannetti2011}%
  \BibitemOpen
  \bibfield  {author} {\bibinfo {author} {\bibfnamefont {V.}~\bibnamefont
  {Giovannetti}}, \bibinfo {author} {\bibfnamefont {S.}~\bibnamefont {Lloyd}},
  \ and\ \bibinfo {author} {\bibfnamefont {L.}~\bibnamefont {Maccone}},\ }\href
  {\doibase 10.1038/nphoton.2011.35} {\bibfield  {journal} {\bibinfo  {journal}
  {Nat. Photonics}\ }\textbf {\bibinfo {volume} {5}},\ \bibinfo {pages} {222}
  (\bibinfo {year} {2011})}\BibitemShut {NoStop}%
\bibitem [{\citenamefont {Lloyd}(2008)}]{Lloyd2008}%
  \BibitemOpen
  \bibfield  {author} {\bibinfo {author} {\bibfnamefont {S.}~\bibnamefont
  {Lloyd}},\ }\href {\doibase 10.1126/science.1160627} {\bibfield  {journal}
  {\bibinfo  {journal} {Science}\ }\textbf {\bibinfo {volume} {321}},\ \bibinfo
  {pages} {1463} (\bibinfo {year} {2008})}\BibitemShut {NoStop}%
\bibitem [{\citenamefont {Benedetti}\ \emph {et~al.}(2019)\citenamefont
  {Benedetti}, \citenamefont {Grant}, \citenamefont {Wossnig},\ and\
  \citenamefont {Severini}}]{Benedetti2019}%
  \BibitemOpen
  \bibfield  {author} {\bibinfo {author} {\bibfnamefont {M.}~\bibnamefont
  {Benedetti}}, \bibinfo {author} {\bibfnamefont {E.}~\bibnamefont {Grant}},
  \bibinfo {author} {\bibfnamefont {L.}~\bibnamefont {Wossnig}}, \ and\
  \bibinfo {author} {\bibfnamefont {S.}~\bibnamefont {Severini}},\ }\href
  {\doibase 10.1088/1367-2630/ab14b5} {\bibfield  {journal} {\bibinfo
  {journal} {New J. Phys.}\ }\textbf {\bibinfo {volume} {21}},\ \bibinfo
  {pages} {043023} (\bibinfo {year} {2019})}\BibitemShut {NoStop}%
\bibitem [{\citenamefont {IBM}(2019)}]{IBM2019}%
  \BibitemOpen
  \bibfield  {author} {\bibinfo {author} {\bibnamefont {IBM}},\ }\href
  {https://www.research.ibm.com/ibm-q/technology/devices/} {\enquote {\bibinfo
  {title} {{Quantum systems - IBM Q}},}\ } (\bibinfo {year} {2019})\BibitemShut
  {NoStop}%
\bibitem [{\citenamefont {Hong}\ \emph {et~al.}(2019)\citenamefont {Hong},
  \citenamefont {Papageorge}, \citenamefont {Sivarajah}, \citenamefont
  {Crossman}, \citenamefont {Dider}, \citenamefont {Polloreno}, \citenamefont
  {Sete}, \citenamefont {Turkowski}, \citenamefont {da~Silva},\ and\
  \citenamefont {Johnson}}]{Hong2019}%
  \BibitemOpen
  \bibfield  {author} {\bibinfo {author} {\bibfnamefont {S.~S.}\ \bibnamefont
  {Hong}}, \bibinfo {author} {\bibfnamefont {A.~T.}\ \bibnamefont
  {Papageorge}}, \bibinfo {author} {\bibfnamefont {P.}~\bibnamefont
  {Sivarajah}}, \bibinfo {author} {\bibfnamefont {G.}~\bibnamefont {Crossman}},
  \bibinfo {author} {\bibfnamefont {N.}~\bibnamefont {Dider}}, \bibinfo
  {author} {\bibfnamefont {A.~M.}\ \bibnamefont {Polloreno}}, \bibinfo {author}
  {\bibfnamefont {E.~A.}\ \bibnamefont {Sete}}, \bibinfo {author}
  {\bibfnamefont {S.~W.}\ \bibnamefont {Turkowski}}, \bibinfo {author}
  {\bibfnamefont {M.~P.}\ \bibnamefont {da~Silva}}, \ and\ \bibinfo {author}
  {\bibfnamefont {B.~R.}\ \bibnamefont {Johnson}},\ }\href
  {http://arxiv.org/abs/1901.08035} {\bibfield  {journal} {\bibinfo  {journal}
  {{Preprint}}\ } (\bibinfo {year} {2019})},\ \Eprint
  {http://arxiv.org/abs/1901.08035} {arXiv:1901.08035} \BibitemShut {NoStop}%
\bibitem [{\citenamefont {Chen}\ \emph {et~al.}(2014)\citenamefont {Chen},
  \citenamefont {Neill}, \citenamefont {Roushan}, \citenamefont {Leung},
  \citenamefont {Fang}, \citenamefont {Barends}, \citenamefont {Kelly},
  \citenamefont {Campbell}, \citenamefont {Chen}, \citenamefont {Chiaro},
  \citenamefont {Dunsworth}, \citenamefont {Jeffrey}, \citenamefont {Megrant},
  \citenamefont {Mutus}, \citenamefont {O'Malley}, \citenamefont {Quintana},
  \citenamefont {Sank}, \citenamefont {Vainsencher}, \citenamefont {Wenner},
  \citenamefont {White}, \citenamefont {Geller}, \citenamefont {Cleland},\ and\
  \citenamefont {Martinis}}]{Chen2014}%
  \BibitemOpen
  \bibfield  {author} {\bibinfo {author} {\bibfnamefont {Y.}~\bibnamefont
  {Chen}}, \bibinfo {author} {\bibfnamefont {C.}~\bibnamefont {Neill}},
  \bibinfo {author} {\bibfnamefont {P.}~\bibnamefont {Roushan}}, \bibinfo
  {author} {\bibfnamefont {N.}~\bibnamefont {Leung}}, \bibinfo {author}
  {\bibfnamefont {M.}~\bibnamefont {Fang}}, \bibinfo {author} {\bibfnamefont
  {R.}~\bibnamefont {Barends}}, \bibinfo {author} {\bibfnamefont
  {J.}~\bibnamefont {Kelly}}, \bibinfo {author} {\bibfnamefont
  {B.}~\bibnamefont {Campbell}}, \bibinfo {author} {\bibfnamefont
  {Z.}~\bibnamefont {Chen}}, \bibinfo {author} {\bibfnamefont {B.}~\bibnamefont
  {Chiaro}}, \bibinfo {author} {\bibfnamefont {A.}~\bibnamefont {Dunsworth}},
  \bibinfo {author} {\bibfnamefont {E.}~\bibnamefont {Jeffrey}}, \bibinfo
  {author} {\bibfnamefont {A.}~\bibnamefont {Megrant}}, \bibinfo {author}
  {\bibfnamefont {J.~Y.}\ \bibnamefont {Mutus}}, \bibinfo {author}
  {\bibfnamefont {P.~J.~J.}\ \bibnamefont {O'Malley}}, \bibinfo {author}
  {\bibfnamefont {C.~M.}\ \bibnamefont {Quintana}}, \bibinfo {author}
  {\bibfnamefont {D.}~\bibnamefont {Sank}}, \bibinfo {author} {\bibfnamefont
  {A.}~\bibnamefont {Vainsencher}}, \bibinfo {author} {\bibfnamefont
  {J.}~\bibnamefont {Wenner}}, \bibinfo {author} {\bibfnamefont {T.~C.}\
  \bibnamefont {White}}, \bibinfo {author} {\bibfnamefont {M.~R.}\ \bibnamefont
  {Geller}}, \bibinfo {author} {\bibfnamefont {A.~N.}\ \bibnamefont {Cleland}},
  \ and\ \bibinfo {author} {\bibfnamefont {J.~M.}\ \bibnamefont {Martinis}},\
  }\href {\doibase 10.1103/PhysRevLett.113.220502} {\bibfield  {journal}
  {\bibinfo  {journal} {Phys. Rev. Lett.}\ }\textbf {\bibinfo {volume} {113}},\
  \bibinfo {pages} {220502} (\bibinfo {year} {2014})}\BibitemShut {NoStop}%
\bibitem [{\citenamefont {Chen}\ \emph {et~al.}(2018)\citenamefont {Chen},
  \citenamefont {Wossnig}, \citenamefont {Severini}, \citenamefont {Neven},\
  and\ \citenamefont {Mohseni}}]{Chen2018}%
  \BibitemOpen
  \bibfield  {author} {\bibinfo {author} {\bibfnamefont {H.}~\bibnamefont
  {Chen}}, \bibinfo {author} {\bibfnamefont {L.}~\bibnamefont {Wossnig}},
  \bibinfo {author} {\bibfnamefont {S.}~\bibnamefont {Severini}}, \bibinfo
  {author} {\bibfnamefont {H.}~\bibnamefont {Neven}}, \ and\ \bibinfo {author}
  {\bibfnamefont {M.}~\bibnamefont {Mohseni}},\ }\href
  {http://arxiv.org/abs/1805.08654} {\bibfield  {journal} {\bibinfo  {journal}
  {{Preprint}}\ } (\bibinfo {year} {2018})},\ \Eprint
  {http://arxiv.org/abs/1805.08654} {arXiv:1805.08654} \BibitemShut {NoStop}%
\bibitem [{\citenamefont {Goodfellow}\ \emph {et~al.}(2016)\citenamefont
  {Goodfellow}, \citenamefont {Bengio},\ and\ \citenamefont
  {Courvillle}}]{Goodfellow2016}%
  \BibitemOpen
  \bibfield  {author} {\bibinfo {author} {\bibfnamefont {I.}~\bibnamefont
  {Goodfellow}}, \bibinfo {author} {\bibfnamefont {Y.}~\bibnamefont {Bengio}},
  \ and\ \bibinfo {author} {\bibfnamefont {A.}~\bibnamefont {Courvillle}},\
  }\href {www.deeplearningbook.org} {\emph {\bibinfo {title} {{Deep
  learning}}}}\ (\bibinfo  {publisher} {MIT Press},\ \bibinfo {year}
  {2016})\BibitemShut {NoStop}%
\bibitem [{\citenamefont {Arjovsky}\ \emph {et~al.}(2015)\citenamefont
  {Arjovsky}, \citenamefont {Shah},\ and\ \citenamefont
  {Bengio}}]{Arjovsky2015}%
  \BibitemOpen
  \bibfield  {author} {\bibinfo {author} {\bibfnamefont {M.}~\bibnamefont
  {Arjovsky}}, \bibinfo {author} {\bibfnamefont {A.}~\bibnamefont {Shah}}, \
  and\ \bibinfo {author} {\bibfnamefont {Y.}~\bibnamefont {Bengio}},\ }\href
  {http://arxiv.org/abs/1511.06464} {\bibfield  {journal} {\bibinfo  {journal}
  {{Preprint}}\ } (\bibinfo {year} {2015})},\ \Eprint
  {http://arxiv.org/abs/1511.06464} {arXiv:1511.06464} \BibitemShut {NoStop}%
\bibitem [{\citenamefont {Mohseni}\ \emph {et~al.}(2004)\citenamefont
  {Mohseni}, \citenamefont {Steinberg},\ and\ \citenamefont
  {Bergou}}]{Mohseni2004}%
  \BibitemOpen
  \bibfield  {author} {\bibinfo {author} {\bibfnamefont {M.}~\bibnamefont
  {Mohseni}}, \bibinfo {author} {\bibfnamefont {A.~M.}\ \bibnamefont
  {Steinberg}}, \ and\ \bibinfo {author} {\bibfnamefont {J.~A.}\ \bibnamefont
  {Bergou}},\ }\href {\doibase 10.1103/PhysRevLett.93.200403} {\bibfield
  {journal} {\bibinfo  {journal} {Phys. Rev. Lett.}\ }\textbf {\bibinfo
  {volume} {93}},\ \bibinfo {pages} {200403} (\bibinfo {year}
  {2004})}\BibitemShut {NoStop}%
\bibitem [{\citenamefont {Srivastava}\ \emph {et~al.}(2014)\citenamefont
  {Srivastava}, \citenamefont {Hinton}, \citenamefont {Krizhevsky},
  \citenamefont {Sutskever},\ and\ \citenamefont
  {Salakhutdinov}}]{Srivastava2014}%
  \BibitemOpen
  \bibfield  {author} {\bibinfo {author} {\bibfnamefont {N.}~\bibnamefont
  {Srivastava}}, \bibinfo {author} {\bibfnamefont {G.}~\bibnamefont {Hinton}},
  \bibinfo {author} {\bibfnamefont {A.}~\bibnamefont {Krizhevsky}}, \bibinfo
  {author} {\bibfnamefont {I.}~\bibnamefont {Sutskever}}, \ and\ \bibinfo
  {author} {\bibfnamefont {R.}~\bibnamefont {Salakhutdinov}},\ }\href
  {http://jmlr.org/papers/v15/srivastava14a.html} {\bibfield  {journal}
  {\bibinfo  {journal} {Journal of Machine Learning Research}\ }\textbf
  {\bibinfo {volume} {15}},\ \bibinfo {pages} {1929} (\bibinfo {year}
  {2014})}\BibitemShut {NoStop}%
\bibitem [{\citenamefont {Negnevitsky}\ \emph {et~al.}(2018)\citenamefont
  {Negnevitsky}, \citenamefont {Marinelli}, \citenamefont {Mehta},
  \citenamefont {Lo}, \citenamefont {Flühmann},\ and\ \citenamefont
  {Home}}]{Negnevitsky2018}%
  \BibitemOpen
  \bibfield  {author} {\bibinfo {author} {\bibfnamefont {V.}~\bibnamefont
  {Negnevitsky}}, \bibinfo {author} {\bibfnamefont {M.}~\bibnamefont
  {Marinelli}}, \bibinfo {author} {\bibfnamefont {K.~K.}\ \bibnamefont
  {Mehta}}, \bibinfo {author} {\bibfnamefont {H.-Y.}\ \bibnamefont {Lo}},
  \bibinfo {author} {\bibfnamefont {C.}~\bibnamefont {Flühmann}}, \ and\
  \bibinfo {author} {\bibfnamefont {J.~P.}\ \bibnamefont {Home}},\ }\href
  {\doibase 10.1038/s41586-018-0668-z} {\bibfield  {journal} {\bibinfo
  {journal} {Nature}\ }\textbf {\bibinfo {volume} {563}},\ \bibinfo {pages}
  {527} (\bibinfo {year} {2018})}\BibitemShut {NoStop}%
\bibitem [{\citenamefont {Kingma}\ and\ \citenamefont {Ba}(2014)}]{Kingma2014}%
  \BibitemOpen
  \bibfield  {author} {\bibinfo {author} {\bibfnamefont {D.~P.}\ \bibnamefont
  {Kingma}}\ and\ \bibinfo {author} {\bibfnamefont {J.}~\bibnamefont {Ba}},\
  }\href {http://arxiv.org/abs/1412.6980} {\bibfield  {journal} {\bibinfo
  {journal} {{Preprint}}\ } (\bibinfo {year} {2014})},\ \Eprint
  {http://arxiv.org/abs/1412.6980} {arXiv:1412.6980} \BibitemShut {NoStop}%
\bibitem [{\citenamefont {Gokhale}\ \emph {et~al.}(2019)\citenamefont
  {Gokhale}, \citenamefont {Ding}, \citenamefont {Propson}, \citenamefont
  {Winkler}, \citenamefont {Leung}, \citenamefont {Shi}, \citenamefont
  {Schuster}, \citenamefont {Hoffmann},\ and\ \citenamefont
  {Chong}}]{Gokhale2019}%
  \BibitemOpen
  \bibfield  {author} {\bibinfo {author} {\bibfnamefont {P.}~\bibnamefont
  {Gokhale}}, \bibinfo {author} {\bibfnamefont {Y.}~\bibnamefont {Ding}},
  \bibinfo {author} {\bibfnamefont {T.}~\bibnamefont {Propson}}, \bibinfo
  {author} {\bibfnamefont {C.}~\bibnamefont {Winkler}}, \bibinfo {author}
  {\bibfnamefont {N.}~\bibnamefont {Leung}}, \bibinfo {author} {\bibfnamefont
  {Y.}~\bibnamefont {Shi}}, \bibinfo {author} {\bibfnamefont {D.~I.}\
  \bibnamefont {Schuster}}, \bibinfo {author} {\bibfnamefont {H.}~\bibnamefont
  {Hoffmann}}, \ and\ \bibinfo {author} {\bibfnamefont {F.~T.}\ \bibnamefont
  {Chong}},\ }in\ \href {\doibase 10.1145/3352460.3358313} {\emph {\bibinfo
  {booktitle} {Proceedings of the 52nd Annual IEEE/ACM International Symposium
  on Microarchitecture - MICRO '52}}}\ (\bibinfo  {publisher} {ACM Press},\
  \bibinfo {address} {New York, New York, USA},\ \bibinfo {year} {2019})\ pp.\
  \bibinfo {pages} {266--278},\ \Eprint {http://arxiv.org/abs/1909.07522}
  {arXiv:1909.07522} \BibitemShut {NoStop}%
\bibitem [{\citenamefont {Sweke}\ \emph {et~al.}(2019)\citenamefont {Sweke},
  \citenamefont {Wilde}, \citenamefont {Meyer}, \citenamefont {Schuld},
  \citenamefont {F{\"{a}}hrmann}, \citenamefont {Meynard-Piganeau},\ and\
  \citenamefont {Eisert}}]{Sweke2019}%
  \BibitemOpen
  \bibfield  {author} {\bibinfo {author} {\bibfnamefont {R.}~\bibnamefont
  {Sweke}}, \bibinfo {author} {\bibfnamefont {F.}~\bibnamefont {Wilde}},
  \bibinfo {author} {\bibfnamefont {J.}~\bibnamefont {Meyer}}, \bibinfo
  {author} {\bibfnamefont {M.}~\bibnamefont {Schuld}}, \bibinfo {author}
  {\bibfnamefont {P.~K.}\ \bibnamefont {F{\"{a}}hrmann}}, \bibinfo {author}
  {\bibfnamefont {B.}~\bibnamefont {Meynard-Piganeau}}, \ and\ \bibinfo
  {author} {\bibfnamefont {J.}~\bibnamefont {Eisert}},\ }\href
  {http://arxiv.org/abs/1910.01155} {\bibfield  {journal} {\bibinfo  {journal}
  {{Preprint}}\ } (\bibinfo {year} {2019})},\ \Eprint
  {http://arxiv.org/abs/1910.01155} {arXiv:1910.01155} \BibitemShut {NoStop}%
\bibitem [{\citenamefont {Hamilton}\ \emph {et~al.}(2019)\citenamefont
  {Hamilton}, \citenamefont {Dumitrescu},\ and\ \citenamefont
  {Pooser}}]{Hamilton2019}%
  \BibitemOpen
  \bibfield  {author} {\bibinfo {author} {\bibfnamefont {K.~E.}\ \bibnamefont
  {Hamilton}}, \bibinfo {author} {\bibfnamefont {E.~F.}\ \bibnamefont
  {Dumitrescu}}, \ and\ \bibinfo {author} {\bibfnamefont {R.~C.}\ \bibnamefont
  {Pooser}},\ }\href {\doibase 10.1103/PhysRevA.99.062323} {\bibfield
  {journal} {\bibinfo  {journal} {Phys. Rev. A}\ }\textbf {\bibinfo {volume}
  {99}},\ \bibinfo {pages} {062323} (\bibinfo {year} {2019})},\ \Eprint
  {http://arxiv.org/abs/1811.09905} {arXiv:1811.09905} \BibitemShut {NoStop}%
\bibitem [{\citenamefont {Liu}\ \emph {et~al.}(2019)\citenamefont {Liu},
  \citenamefont {Zhang}, \citenamefont {Wan},\ and\ \citenamefont
  {Wang}}]{Liu2019}%
  \BibitemOpen
  \bibfield  {author} {\bibinfo {author} {\bibfnamefont {J.-G.}\ \bibnamefont
  {Liu}}, \bibinfo {author} {\bibfnamefont {Y.-H.}\ \bibnamefont {Zhang}},
  \bibinfo {author} {\bibfnamefont {Y.}~\bibnamefont {Wan}}, \ and\ \bibinfo
  {author} {\bibfnamefont {L.}~\bibnamefont {Wang}},\ }\href {\doibase
  10.1103/PhysRevResearch.1.023025} {\bibfield  {journal} {\bibinfo  {journal}
  {Phys. Rev. Research}\ }\textbf {\bibinfo {volume} {1}},\ \bibinfo {pages}
  {023025} (\bibinfo {year} {2019})},\ \Eprint
  {http://arxiv.org/abs/1902.02663} {arXiv:1902.02663} \BibitemShut {NoStop}%
\bibitem [{\citenamefont {Neelakantan}\ \emph {et~al.}(2015)\citenamefont
  {Neelakantan}, \citenamefont {Vilnis}, \citenamefont {Le}, \citenamefont
  {Sutskever}, \citenamefont {Kaiser}, \citenamefont {Kurach},\ and\
  \citenamefont {Martens}}]{Neelakantan2015}%
  \BibitemOpen
  \bibfield  {author} {\bibinfo {author} {\bibfnamefont {A.}~\bibnamefont
  {Neelakantan}}, \bibinfo {author} {\bibfnamefont {L.}~\bibnamefont {Vilnis}},
  \bibinfo {author} {\bibfnamefont {Q.~V.}\ \bibnamefont {Le}}, \bibinfo
  {author} {\bibfnamefont {I.}~\bibnamefont {Sutskever}}, \bibinfo {author}
  {\bibfnamefont {L.}~\bibnamefont {Kaiser}}, \bibinfo {author} {\bibfnamefont
  {K.}~\bibnamefont {Kurach}}, \ and\ \bibinfo {author} {\bibfnamefont
  {J.}~\bibnamefont {Martens}},\ }\href {http://arxiv.org/abs/1511.06807}
  {\bibfield  {journal} {\bibinfo  {journal} {{Preprint}}\ } (\bibinfo {year}
  {2015})},\ \Eprint {http://arxiv.org/abs/1511.06807} {arXiv:1511.06807}
  \BibitemShut {NoStop}%
\bibitem [{\citenamefont {Ostaszewski}\ \emph {et~al.}(2019)\citenamefont
  {Ostaszewski}, \citenamefont {Grant},\ and\ \citenamefont
  {Benedetti}}]{Ostaszewski2019}%
  \BibitemOpen
  \bibfield  {author} {\bibinfo {author} {\bibfnamefont {M.}~\bibnamefont
  {Ostaszewski}}, \bibinfo {author} {\bibfnamefont {E.}~\bibnamefont {Grant}},
  \ and\ \bibinfo {author} {\bibfnamefont {M.}~\bibnamefont {Benedetti}},\
  }\href {http://arxiv.org/abs/1905.09692} {\bibfield  {journal} {\bibinfo
  {journal} {{Preprint}}\ } (\bibinfo {year} {2019})},\ \Eprint
  {http://arxiv.org/abs/1905.09692} {arXiv:1905.09692} \BibitemShut {NoStop}%
\bibitem [{\citenamefont {McClean}\ \emph {et~al.}(2018)\citenamefont
  {McClean}, \citenamefont {Boixo}, \citenamefont {Smelyanskiy}, \citenamefont
  {Babbush},\ and\ \citenamefont {Neven}}]{McClean2018}%
  \BibitemOpen
  \bibfield  {author} {\bibinfo {author} {\bibfnamefont {J.~R.}\ \bibnamefont
  {McClean}}, \bibinfo {author} {\bibfnamefont {S.}~\bibnamefont {Boixo}},
  \bibinfo {author} {\bibfnamefont {V.~N.}\ \bibnamefont {Smelyanskiy}},
  \bibinfo {author} {\bibfnamefont {R.}~\bibnamefont {Babbush}}, \ and\
  \bibinfo {author} {\bibfnamefont {H.}~\bibnamefont {Neven}},\ }\href
  {\doibase 10.1038/s41467-018-07090-4} {\bibfield  {journal} {\bibinfo
  {journal} {Nat. Commun.}\ }\textbf {\bibinfo {volume} {9}},\ \bibinfo {pages}
  {4812} (\bibinfo {year} {2018})}\BibitemShut {NoStop}%
\bibitem [{\citenamefont {Nelder}\ and\ \citenamefont
  {Mead}(1965)}]{Nelder1965}%
  \BibitemOpen
  \bibfield  {author} {\bibinfo {author} {\bibfnamefont {J.~A.}\ \bibnamefont
  {Nelder}}\ and\ \bibinfo {author} {\bibfnamefont {R.}~\bibnamefont {Mead}},\
  }\href {\doibase 10.1093/comjnl/7.4.308} {\bibfield  {journal} {\bibinfo
  {journal} {The Computer Journal}\ }\textbf {\bibinfo {volume} {7}},\ \bibinfo
  {pages} {308} (\bibinfo {year} {1965})}\BibitemShut {NoStop}%
\bibitem [{\citenamefont {Mitarai}\ \emph {et~al.}(2018)\citenamefont
  {Mitarai}, \citenamefont {Negoro}, \citenamefont {Kitagawa},\ and\
  \citenamefont {Fujii}}]{Mitarai2018}%
  \BibitemOpen
  \bibfield  {author} {\bibinfo {author} {\bibfnamefont {K.}~\bibnamefont
  {Mitarai}}, \bibinfo {author} {\bibfnamefont {M.}~\bibnamefont {Negoro}},
  \bibinfo {author} {\bibfnamefont {M.}~\bibnamefont {Kitagawa}}, \ and\
  \bibinfo {author} {\bibfnamefont {K.}~\bibnamefont {Fujii}},\ }\href
  {\doibase 10.1103/PhysRevA.98.032309} {\bibfield  {journal} {\bibinfo
  {journal} {Phys. Rev. A}\ }\textbf {\bibinfo {volume} {98}},\ \bibinfo
  {pages} {032309} (\bibinfo {year} {2018})},\ \Eprint
  {http://arxiv.org/abs/1803.00745} {arXiv:1803.00745} \BibitemShut {NoStop}%
\bibitem [{\citenamefont {Schuld}\ \emph {et~al.}(2019)\citenamefont {Schuld},
  \citenamefont {Bergholm}, \citenamefont {Gogolin}, \citenamefont {Izaac},\
  and\ \citenamefont {Killoran}}]{Schuld2018}%
  \BibitemOpen
  \bibfield  {author} {\bibinfo {author} {\bibfnamefont {M.}~\bibnamefont
  {Schuld}}, \bibinfo {author} {\bibfnamefont {V.}~\bibnamefont {Bergholm}},
  \bibinfo {author} {\bibfnamefont {C.}~\bibnamefont {Gogolin}}, \bibinfo
  {author} {\bibfnamefont {J.}~\bibnamefont {Izaac}}, \ and\ \bibinfo {author}
  {\bibfnamefont {N.}~\bibnamefont {Killoran}},\ }\href {\doibase
  10.1103/PhysRevA.99.032331} {\bibfield  {journal} {\bibinfo  {journal} {Phys.
  Rev. A}\ }\textbf {\bibinfo {volume} {99}},\ \bibinfo {pages} {032331}
  (\bibinfo {year} {2019})},\ \Eprint {http://arxiv.org/abs/1811.11184v1}
  {arXiv:1811.11184v1} \BibitemShut {NoStop}%
\bibitem [{\citenamefont {Parrish}\ \emph {et~al.}(2019)\citenamefont
  {Parrish}, \citenamefont {Hohenstein}, \citenamefont {McMahon},\ and\
  \citenamefont {Martinez}}]{Parrish2019}%
  \BibitemOpen
  \bibfield  {author} {\bibinfo {author} {\bibfnamefont {R.~M.}\ \bibnamefont
  {Parrish}}, \bibinfo {author} {\bibfnamefont {E.~G.}\ \bibnamefont
  {Hohenstein}}, \bibinfo {author} {\bibfnamefont {P.~L.}\ \bibnamefont
  {McMahon}}, \ and\ \bibinfo {author} {\bibfnamefont {T.~J.}\ \bibnamefont
  {Martinez}},\ }\href {http://arxiv.org/abs/1906.08728} {\bibfield  {journal}
  {\bibinfo  {journal} {{Preprint}}\ } (\bibinfo {year} {2019})},\ \Eprint
  {http://arxiv.org/abs/1906.08728} {arXiv:1906.08728} \BibitemShut {NoStop}%
\bibitem [{\citenamefont {Peruzzo}\ \emph {et~al.}(2014)\citenamefont
  {Peruzzo}, \citenamefont {McClean}, \citenamefont {Shadbolt}, \citenamefont
  {Yung}, \citenamefont {Zhou}, \citenamefont {Love}, \citenamefont
  {Aspuru-Guzik},\ and\ \citenamefont {O'Brien}}]{Peruzzo2014}%
  \BibitemOpen
  \bibfield  {author} {\bibinfo {author} {\bibfnamefont {A.}~\bibnamefont
  {Peruzzo}}, \bibinfo {author} {\bibfnamefont {J.}~\bibnamefont {McClean}},
  \bibinfo {author} {\bibfnamefont {P.}~\bibnamefont {Shadbolt}}, \bibinfo
  {author} {\bibfnamefont {M.-H.}\ \bibnamefont {Yung}}, \bibinfo {author}
  {\bibfnamefont {X.-Q.}\ \bibnamefont {Zhou}}, \bibinfo {author}
  {\bibfnamefont {P.~J.}\ \bibnamefont {Love}}, \bibinfo {author}
  {\bibfnamefont {A.}~\bibnamefont {Aspuru-Guzik}}, \ and\ \bibinfo {author}
  {\bibfnamefont {J.~L.}\ \bibnamefont {O'Brien}},\ }\href {\doibase
  10.1038/ncomms5213} {\bibfield  {journal} {\bibinfo  {journal} {Nat.
  Commun.}\ }\textbf {\bibinfo {volume} {5}},\ \bibinfo {pages} {4213}
  (\bibinfo {year} {2014})}\BibitemShut {NoStop}%
\bibitem [{\citenamefont {Nielsen}\ and\ \citenamefont
  {Chuang}(2011)}]{Nielsen2000a}%
  \BibitemOpen
  \bibfield  {author} {\bibinfo {author} {\bibfnamefont {M.~A.}\ \bibnamefont
  {Nielsen}}\ and\ \bibinfo {author} {\bibfnamefont {I.~L.}\ \bibnamefont
  {Chuang}},\ }\href@noop {} {\emph {\bibinfo {title} {{Quantum Computation and
  Quantum Information: 10th Anniversary Edition}}}},\ \bibinfo {edition}
  {10th}\ ed.\ (\bibinfo  {publisher} {Cambridge University Press},\ \bibinfo
  {address} {New York, NY, USA},\ \bibinfo {year} {2011})\BibitemShut {NoStop}%
\bibitem [{\citenamefont {Sim}\ \emph {et~al.}(2019)\citenamefont {Sim},
  \citenamefont {Johnson},\ and\ \citenamefont {Aspuru-Guzik}}]{Sim2019}%
  \BibitemOpen
  \bibfield  {author} {\bibinfo {author} {\bibfnamefont {S.}~\bibnamefont
  {Sim}}, \bibinfo {author} {\bibfnamefont {P.~D.}\ \bibnamefont {Johnson}}, \
  and\ \bibinfo {author} {\bibfnamefont {A.}~\bibnamefont {Aspuru-Guzik}},\
  }\href {http://arxiv.org/abs/1905.10876} {\bibfield  {journal} {\bibinfo
  {journal} {{Preprint}}\ } (\bibinfo {year} {2019})},\ \Eprint
  {http://arxiv.org/abs/1905.10876} {arXiv:1905.10876} \BibitemShut {NoStop}%
\bibitem [{\citenamefont {Knill}(2005)}]{Knill2004}%
  \BibitemOpen
  \bibfield  {author} {\bibinfo {author} {\bibfnamefont {E.}~\bibnamefont
  {Knill}},\ }\href {\doibase 10.1038/nature03350} {\bibfield  {journal}
  {\bibinfo  {journal} {Nature}\ }\textbf {\bibinfo {volume} {434}},\ \bibinfo
  {pages} {39} (\bibinfo {year} {2005})},\ \Eprint
  {http://arxiv.org/abs/0410199} {arXiv:0410199 [quant-ph]} \BibitemShut
  {NoStop}%
\bibitem [{\citenamefont {Gottesman}(2010)}]{Gottesman2010}%
  \BibitemOpen
  \bibfield  {author} {\bibinfo {author} {\bibfnamefont {D.}~\bibnamefont
  {Gottesman}}\ }(\bibinfo {year} {2010})\ pp.\ \bibinfo {pages} {13--58},\
  \Eprint {http://arxiv.org/abs/0904.2557} {arXiv:0904.2557} \BibitemShut
  {NoStop}%
\bibitem [{\citenamefont {McClean}\ \emph {et~al.}(2016)\citenamefont
  {McClean}, \citenamefont {Romero}, \citenamefont {Babbush},\ and\
  \citenamefont {Aspuru-Guzik}}]{Mcclean2016}%
  \BibitemOpen
  \bibfield  {author} {\bibinfo {author} {\bibfnamefont {J.~R.}\ \bibnamefont
  {McClean}}, \bibinfo {author} {\bibfnamefont {J.}~\bibnamefont {Romero}},
  \bibinfo {author} {\bibfnamefont {R.}~\bibnamefont {Babbush}}, \ and\
  \bibinfo {author} {\bibfnamefont {A.}~\bibnamefont {Aspuru-Guzik}},\ }\href
  {\doibase 10.1088/1367-2630/18/2/023023} {\bibfield  {journal} {\bibinfo
  {journal} {New J. Phys.}\ }\textbf {\bibinfo {volume} {18}},\ \bibinfo
  {pages} {023023} (\bibinfo {year} {2016})},\ \Eprint
  {http://arxiv.org/abs/1509.04279v1} {arXiv:1509.04279v1} \BibitemShut
  {NoStop}%
\bibitem [{\citenamefont {Abadi}\ \emph {et~al.}(2016)\citenamefont {Abadi},
  \citenamefont {Agarwal}, \citenamefont {Barham}, \citenamefont {Brevdo},
  \citenamefont {Chen}, \citenamefont {Citro}, \citenamefont {Corrado},
  \citenamefont {Davis}, \citenamefont {Dean}, \citenamefont {Devin},
  \citenamefont {Ghemawat}, \citenamefont {Goodfellow}, \citenamefont {Harp},
  \citenamefont {Irving}, \citenamefont {Isard}, \citenamefont {Jia},
  \citenamefont {Jozefowicz}, \citenamefont {Kaiser}, \citenamefont {Kudlur},
  \citenamefont {Levenberg}, \citenamefont {Mane}, \citenamefont {Monga},
  \citenamefont {Moore}, \citenamefont {Murray}, \citenamefont {Olah},
  \citenamefont {Schuster}, \citenamefont {Shlens}, \citenamefont {Steiner},
  \citenamefont {Sutskever}, \citenamefont {Talwar}, \citenamefont {Tucker},
  \citenamefont {Vanhoucke}, \citenamefont {Vasudevan}, \citenamefont {Viegas},
  \citenamefont {Vinyals}, \citenamefont {Warden}, \citenamefont {Wattenberg},
  \citenamefont {Wicke}, \citenamefont {Yu},\ and\ \citenamefont
  {Zheng}}]{Abadi2016a}%
  \BibitemOpen
  \bibfield  {author} {\bibinfo {author} {\bibfnamefont {M.}~\bibnamefont
  {Abadi}}, \bibinfo {author} {\bibfnamefont {A.}~\bibnamefont {Agarwal}},
  \bibinfo {author} {\bibfnamefont {P.}~\bibnamefont {Barham}}, \bibinfo
  {author} {\bibfnamefont {E.}~\bibnamefont {Brevdo}}, \bibinfo {author}
  {\bibfnamefont {Z.}~\bibnamefont {Chen}}, \bibinfo {author} {\bibfnamefont
  {C.}~\bibnamefont {Citro}}, \bibinfo {author} {\bibfnamefont {G.~S.}\
  \bibnamefont {Corrado}}, \bibinfo {author} {\bibfnamefont {A.}~\bibnamefont
  {Davis}}, \bibinfo {author} {\bibfnamefont {J.}~\bibnamefont {Dean}},
  \bibinfo {author} {\bibfnamefont {M.}~\bibnamefont {Devin}}, \bibinfo
  {author} {\bibfnamefont {S.}~\bibnamefont {Ghemawat}}, \bibinfo {author}
  {\bibfnamefont {I.}~\bibnamefont {Goodfellow}}, \bibinfo {author}
  {\bibfnamefont {A.}~\bibnamefont {Harp}}, \bibinfo {author} {\bibfnamefont
  {G.}~\bibnamefont {Irving}}, \bibinfo {author} {\bibfnamefont
  {M.}~\bibnamefont {Isard}}, \bibinfo {author} {\bibfnamefont
  {Y.}~\bibnamefont {Jia}}, \bibinfo {author} {\bibfnamefont {R.}~\bibnamefont
  {Jozefowicz}}, \bibinfo {author} {\bibfnamefont {L.}~\bibnamefont {Kaiser}},
  \bibinfo {author} {\bibfnamefont {M.}~\bibnamefont {Kudlur}}, \bibinfo
  {author} {\bibfnamefont {J.}~\bibnamefont {Levenberg}}, \bibinfo {author}
  {\bibfnamefont {D.}~\bibnamefont {Mane}}, \bibinfo {author} {\bibfnamefont
  {R.}~\bibnamefont {Monga}}, \bibinfo {author} {\bibfnamefont
  {S.}~\bibnamefont {Moore}}, \bibinfo {author} {\bibfnamefont
  {D.}~\bibnamefont {Murray}}, \bibinfo {author} {\bibfnamefont
  {C.}~\bibnamefont {Olah}}, \bibinfo {author} {\bibfnamefont {M.}~\bibnamefont
  {Schuster}}, \bibinfo {author} {\bibfnamefont {J.}~\bibnamefont {Shlens}},
  \bibinfo {author} {\bibfnamefont {B.}~\bibnamefont {Steiner}}, \bibinfo
  {author} {\bibfnamefont {I.}~\bibnamefont {Sutskever}}, \bibinfo {author}
  {\bibfnamefont {K.}~\bibnamefont {Talwar}}, \bibinfo {author} {\bibfnamefont
  {P.}~\bibnamefont {Tucker}}, \bibinfo {author} {\bibfnamefont
  {V.}~\bibnamefont {Vanhoucke}}, \bibinfo {author} {\bibfnamefont
  {V.}~\bibnamefont {Vasudevan}}, \bibinfo {author} {\bibfnamefont
  {F.}~\bibnamefont {Viegas}}, \bibinfo {author} {\bibfnamefont
  {O.}~\bibnamefont {Vinyals}}, \bibinfo {author} {\bibfnamefont
  {P.}~\bibnamefont {Warden}}, \bibinfo {author} {\bibfnamefont
  {M.}~\bibnamefont {Wattenberg}}, \bibinfo {author} {\bibfnamefont
  {M.}~\bibnamefont {Wicke}}, \bibinfo {author} {\bibfnamefont
  {Y.}~\bibnamefont {Yu}}, \ and\ \bibinfo {author} {\bibfnamefont
  {X.}~\bibnamefont {Zheng}},\ }\href {http://arxiv.org/abs/1603.04467}
  {\bibfield  {journal} {\bibinfo  {journal} {{Preprint}}\ } (\bibinfo {year}
  {2016})},\ \Eprint {http://arxiv.org/abs/1603.04467} {arXiv:1603.04467}
  \BibitemShut {NoStop}%
\bibitem [{\citenamefont {Gidney}\ \emph {et~al.}(2018)\citenamefont {Gidney},
  \citenamefont {Bacon},\ and\ \citenamefont {{The Cirq
  Developers}}}]{Gidney2018}%
  \BibitemOpen
  \bibfield  {author} {\bibinfo {author} {\bibfnamefont {C.}~\bibnamefont
  {Gidney}}, \bibinfo {author} {\bibfnamefont {D.}~\bibnamefont {Bacon}}, \
  and\ \bibinfo {author} {\bibnamefont {{The Cirq Developers}}},\ }\href
  {https://github.com/quantumlib/Cirq} {\enquote {\bibinfo {title}
  {{quantumlib/Cirq: A python framework for creating, editing, and invoking
  Noisy Intermediate Scale Quantum (NISQ) circuits.}}}\ } (\bibinfo {year}
  {2018})\BibitemShut {NoStop}%
\bibitem [{\citenamefont {Jozsa}(1994)}]{Jozsa94}%
  \BibitemOpen
  \bibfield  {author} {\bibinfo {author} {\bibfnamefont {R.}~\bibnamefont
  {Jozsa}},\ }\href {\doibase 10.1080/09500349414552171} {\bibfield  {journal}
  {\bibinfo  {journal} {Journal of Modern Optics}\ }\textbf {\bibinfo {volume}
  {41}},\ \bibinfo {pages} {2315} (\bibinfo {year} {1994})},\ \Eprint
  {http://arxiv.org/abs/https://doi.org/10.1080/09500349414552171}
  {https://doi.org/10.1080/09500349414552171} \BibitemShut {NoStop}%
\bibitem [{\citenamefont {Barnett}\ and\ \citenamefont
  {Croke}(2009)}]{Barnett09}%
  \BibitemOpen
  \bibfield  {author} {\bibinfo {author} {\bibfnamefont {S.~M.}\ \bibnamefont
  {Barnett}}\ and\ \bibinfo {author} {\bibfnamefont {S.}~\bibnamefont
  {Croke}},\ }\href {\doibase 10.1364/AOP.1.000238} {\bibfield  {journal}
  {\bibinfo  {journal} {Advances in Optics and Photonics}\ }\textbf {\bibinfo
  {volume} {1}},\ \bibinfo {pages} {238} (\bibinfo {year} {2009})},\ \Eprint
  {http://arxiv.org/abs/0810.1970} {arXiv:0810.1970} \BibitemShut {NoStop}%
\bibitem [{\citenamefont {Bae}\ and\ \citenamefont {Kwek}(2015)}]{Bae17}%
  \BibitemOpen
  \bibfield  {author} {\bibinfo {author} {\bibfnamefont {J.}~\bibnamefont
  {Bae}}\ and\ \bibinfo {author} {\bibfnamefont {L.-C.}\ \bibnamefont {Kwek}},\
  }\href {\doibase 10.1088/1751-8113/48/8/083001} {\bibfield  {journal}
  {\bibinfo  {journal} {Journal of Physics A: Mathematical and Theoretical}\
  }\textbf {\bibinfo {volume} {48}},\ \bibinfo {pages} {083001} (\bibinfo
  {year} {2015})},\ \Eprint {http://arxiv.org/abs/1707.02571}
  {arXiv:1707.02571} \BibitemShut {NoStop}%
\end{thebibliography}%

\end{document}